\documentclass[print]{aa} 
\usepackage[colorlinks=true, linkcolor=blue, citecolor=blue, filecolor=blue, urlcolor=blue]{hyperref}
\usepackage{txfonts}
\usepackage{adjustbox}
\usepackage{indentfirst}
\usepackage{siunitx}
\usepackage{subcaption} 
\usepackage{float}
\usepackage{natbib}

\begin{document} 

\title{Gaia20dsk: A new MNor discovered by the GLORIOUS collaboration}

\subtitle{}

\author{Patrik E. N\'emeth\inst{1,2}
        \and
            Fernando Cruz-Sáenz de Miera\inst{1,3}
        \and
          \'Agnes K\'osp\'al \inst{1,2,4}
        \and
            Eleonora Fiorellino \inst{5,6}
        \and
            Zs\'ofia Nagy \inst{1}
        \and
            M\'at\'e Szil\'agyi \inst{1}
        \and
            Michal Siwak \inst{9}
        \and
            Foteini Lykou \inst{1}
        \and
            M\'aria Kun \inst{1}
        \and
            P\'eter \'Abrah\'am \inst{1}
        \and
            Zs\'ofia Marianna Szab\'o \inst{1,7,8}
        \and 
            Teresa Giannini \inst{10}
        \and
            Lukasz Wyrzykowski \inst{11,12}
          }

\institute{
Konkoly Observatory, HUN-REN Research Centre for Astronomy and Earth Sciences, MTA Centre of Excellence, Konkoly-Thege Mikl\'os \'ut 15-17, 1121 Budapest, Hungary
\and
Institute of Physics and Astronomy, ELTE E\"otv\"os Lor\'and University, Pázmány Péter sétány 1/A, 1117 Budapest, Hungary
         \and
Institut de Recherche en Astrophysique et Planétologie, Université de Toulouse, UT3-PS, OMP, CNRS, 9 av. du Colonel-Roche, 31028 Toulouse Cedex 4, France
         \and
Max Planck Institute for Astronomy, K\"onigstuhl 17, 69117 Heidelberg, Germany
        \and
Instituto de Astrofísica de Canarias, IAC, Vía Láctea s/n, 38205 La Laguna (S.C.Tenerife), Spain
    \and
Departamento de Astrofísica, Universidad de La Laguna, 38206 La Laguna (S.C.Tenerife), Spain
    \and
Scottish Universities Physics Alliance (SUPA), School of Physics and Astronomy, University of St Andrews, North Haugh, St Andrews, KY16 9SS, UK 
    \and
Max-Planck-Institut für Radioastronomie, Auf dem Hügel 69, 53121 Bonn, Germany
    \and    
Mt. Suhora Astronomical Observatory, University of the National Education Commission, ul. Podchorazych 2, 30-084 Krakow, Poland
    \and
INAF—Osservatorio Astronomico di Roma, Via di Frascati, 33, 00078 Monte Porzio Catone, Italy
    \and
ASI, Italian Space Agency, Vi del Politecnico snc, 00133 Rome, Italy
    \and
Warsaw University Observatory, al. Ujazdowskie 4, 00-001 Warsaw
Astrophysics Division, National Centre for Nuclear Research, Pasteura 7, 02-093, Warsaw, Poland    
}
\date{}

    \abstract
   {Among known young stellar objects (YSOs), those exhibiting the most dramatic increases in brightness due to sudden increase in mass accretion rate are eruptive young stars. Gaia20dsk is one of the Gaia-alerted young star candidates that has displayed a double, nonperiodic brightening  resembling that of  other young eruptive stars.}
   {The goal of this work is to determine the physical and accretion properties of Gaia20dsk to confirm its classification as an eruptive young star.}
   {We combined publicly available optical and near-infrared (NIR) photometry with our X-SHOOTER optical/NIR spectrum. In our analysis, we examined the optical and IR light curves from the bursts, reviewing the color-magnitude diagrams across different bands, reporting the detection of emission lines, and providing estimates of the star's accretion rates during the burst.}
   {The optical light curve shows two major and one brief brightening events with an maximum amplitude of $\sim$ 1.8 mag in the last five years. 
   A classification based on spectral index indicates that Gaia20dsk is a flat-spectrum star.
   The X-SHOOTER spectrum exhibit emission lines characteristic of accreting low-to-intermediate-mass young stars, displaying features typical of MNor-type objects. The mass accretion rate is between $\rm (0.5-1.8)\times10^{-6}\,M_\odot/yr$.}
    {Gaia20dsk is an eruptive YSO that exhibits photometric features similar to those of MNors, including its characteristic brightening amplitude and burst duration, along with similar spectroscopic features and accretion rates.}

\titlerunning{Gaia20dsk: a new MNor toward NGC6334 }
\authorrunning{Patrik E. Németh et al.} 

   \maketitle

\section{Introduction}
Young stellar objects (YSOs) display photometric variability in both optical and infrared (IR) bands over timescales ranging from minutes to centuries \citep{Carpenter_2001,Siwak_2018,Herbst94,Kospal12,Cody2014}. 
This variability can be attributed to fluctuations in accretion rates, changes in line-of-sight extinction, or the rotation of accretion hot or cold spots \citep[for a review, see][]{Fischer}.

Among the YSOs, the eruptive young stars exhibit the largest increases in brightness.
These events are caused by the sudden increase in the mass accretion rate.
Historically, these stars have been separated into two classes based on the extent of their brightness variations and the duration of the events.
EX~Lupi-type objects (EXors) undergo outbursts of two to four magnitudes that last from a few months to a year and their mass delivery mechanism is consistent with magnetospheric accretion, albeit at a higher intensity \citep{Jurdana}.
The spectra of EXors closely resemble those of Class II pre-main sequence stars or classical T Tauri stars (CTTS), which are dominated by emission lines.
Their typical mass accretion rate during the outburst is on the order of $10^{-8}-10^{-7} \text{M}_\odot\text{yr}^{-1}$.
In contrast, FU Orionis-type stars (FUors) experience brightenings of 2.5 to 6 magnitudes, with the rise to peak brightness taking from several months to years \citep{Herbig_1977}. 
Their typical mass accretion rate during the outburst is on the order of $10^{-5}-10^{-4} \text{M}_\odot\text{yr}^{-1}$.
The spectral types of FUors vary by wavelength: F--G types are observed in the optical, while K--M types dominate in the near-IR (NIR, \citealt{Hartmann96, Audard}).
This wavelength-dependent spectral type is well modeled by a steady-state viscous accretion disk \citep[e.g.,][]{Liu2022}.

The traditional distinction between FUors and EXors has become increasingly blurred with the discovery of eruptive young stars that do not fit neatly into either category. A growing number of these objects exhibit intermediate properties in terms of outburst duration, mass accretion rate, and spectroscopic features; examples include V1647 Ori \citep{1647}, WISE~1422$-$6115 \citep{Lucas_2020}, and PTF14jg \citep{Hillenbrand_2019b}. 
These stars have been referred to by various names in the literature and one of these classes is MNors \citep{MNOR2017}.
We note that this classification is the same as the one referring to peculiar stars in \citet{peculiar} and  V1647 Ori–like objects in \citet{Fischer}.

Despite these well-known examples, only a few dozen eruptive YSOs have been identified so far.
Their numbers have increased thanks to large-scale surveys and transient detection programs, such as the Gaia alert system, which monitored the sky for sudden brightness variations as part of its
operations, including accretion outbursts.
The Gaia Photometric Science Alerts system announced a list of objects that experienced brightening or dimming in the Gaia $G$ photometric band \citep{Hodgkin}.
When a source experienced a brightening or fading of at least 0.15 magnitudes or a deviation that was at least six times the standard deviation of the baseline flux, an alert was generated and made publicly available\footnote{\url{https://gsaweb.ast.cam.ac.uk/alerts}}.
Several eruptive YSOs have been discovered through this program, including 
Gaia18dvy \citep{Szegedi}, 
Gaia19fct \citep{Park22},
Gaia20eae \citep{Fernando},
Gaia21elv \citep{Zsofi23},
Gaia21bty \citep{Siwak},
Gaia20bdk \citep{Siwak24},
Gaia23bab \citep{Giannini24}, and 
Gaia18cjb \citep{18cjb}.
These sources have been studied and published by members of the  Gaia Science Alerts to Find Eruptive Young Stellar Objects (GLORIOUS) collaboration, introduced here for the first time.

In this paper, we propose Gaia20dsk as another outbursting young star.
The paper is structured as follows.
In Section~\ref{sec:prop}, we describe the properties of Gaia20dsk.
In Section~\ref{sec:data_reduction}, we present the data reduction of the spectrum and its results.
In Section~\ref{sec:szintetikus}, we examine the spectral energy distribution (SED) and Gaia20dsk behavior based on the $JHK_{\rm s}$ color--color diagram, using this information to  classify our source. 
Section~\ref{sec:anal} details our analyses, including the determination of physical properties, calculation of the mass accretion rate, and stellar parameters.
Finally, in Section~\ref{sec:discussion}, we discuss our findings.

\section{Properties of Gaia20dsk}\label{sec:prop}

\subsection{Location and distance}
\begin{figure*}[!htb]
    \centering
    \includegraphics[width=\textwidth]{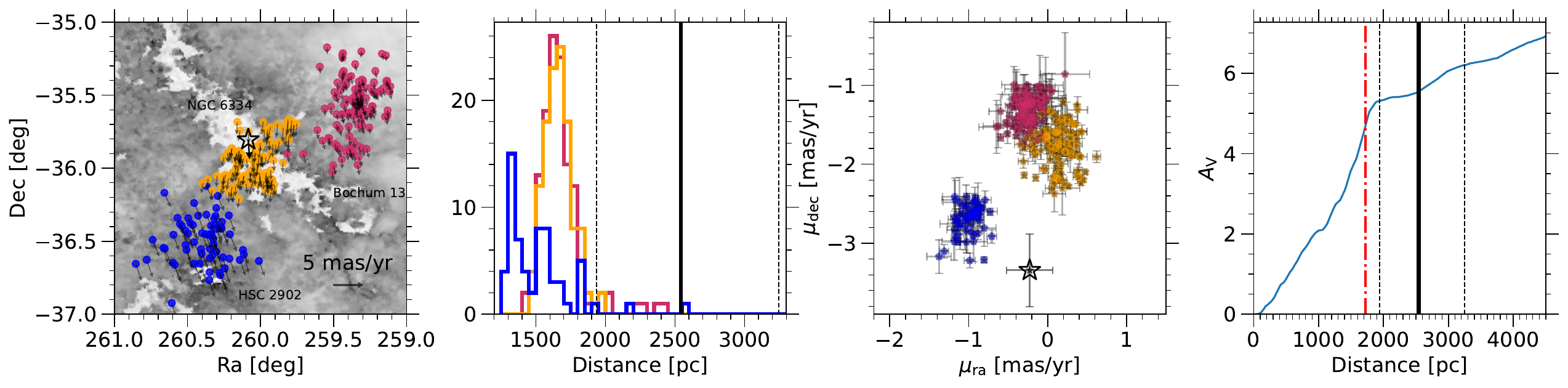}
    \caption{Far left: Spatial distribution of clusters from \citet{Hunt} in the vicinity of Gaia20dsk (shown with a star), overlaid on the DECaPS dust map \citep{Zucker2025}. The arrows indicate the proper motions of the cluster members. Left center: Distance distribution of the cluster members. The thick solid and dashed lines show the $d_{\rm BJ}$ distance of Gaia20dsk and its uncertainties. Right center: Proper motion distribution of the cluster members and Gaia20dsk, shown with error bars. Far right: Cumulative reddening as a function of distance toward Gaia20dsk, based on the data from \citet{Zucker2025}. The red dash-dotted line indicates the most likely distance to Gaia20dsk, while the black solid and dashed lines represent $d_{\rm BJ}$ and its uncertainties.}
    \label{fig:clusters}
\end{figure*}

The coordinates of Gaia20dsk are $ \mathrm{\alpha}_{\rm J2000}~=~17^\mathrm{h}~20^\mathrm{m}~20.13^\mathrm{s}$ and $\mathrm{\delta}_{\rm J2000}~=~ -35^\circ~48\arcmin~17.78\arcsec$, placing it within NGC6334 (see left panel of Fig.~\ref{fig:clusters}), which is a well-known star-forming region \citep{NGC6334}.
Gaia20dsk shares a similar proper motion in right ascension with other members of the group, but with a significantly larger proper motion in declination (see the third panel from the left of Fig.~\ref{fig:clusters}), suggesting that it is not a close member of this cluster.

To constrain the distance to Gaia20dsk, we used the 3D DECaPS dust map \citep{Zucker2025}. 
The cumulative extinction along the line of sight shows a steep rise near 1700 pc (see the last panel of Fig.~\ref{fig:clusters}), which is consistent with the NGC6334 cluster distance of $1650_{-7}^{+14}$\,pc \citep{Hunt}.
Additionally, \citet{Bailer} reported a distance to Gaia20dsk of $2542^{+704}_{-606}$\,pc. 
The estimate relies on photogeometric methods, utilizing Gaia parallax measurements along with Gaia $G$ magnitude and $G_{\rm BP}-G_{\rm RP}$ color data from Gaia EDR3.
Owing to the embedded nature of YSOs, the use of the parallax of a single object to estimate its distance is highly uncertain.
In the following, we describe how we used both the distance to NGC~6334 ($d_{\rm NGC}$) from \cite{Hunt} and the \cite{Bailer} distance ($d_{\rm BJ}$) to calculate the physical parameters of Gaia20dsk and its accretion.

\subsection{Light curve}\label{sec:lc}
\begin{figure}[!htb]
    \centering
    \includegraphics[width=1\linewidth]{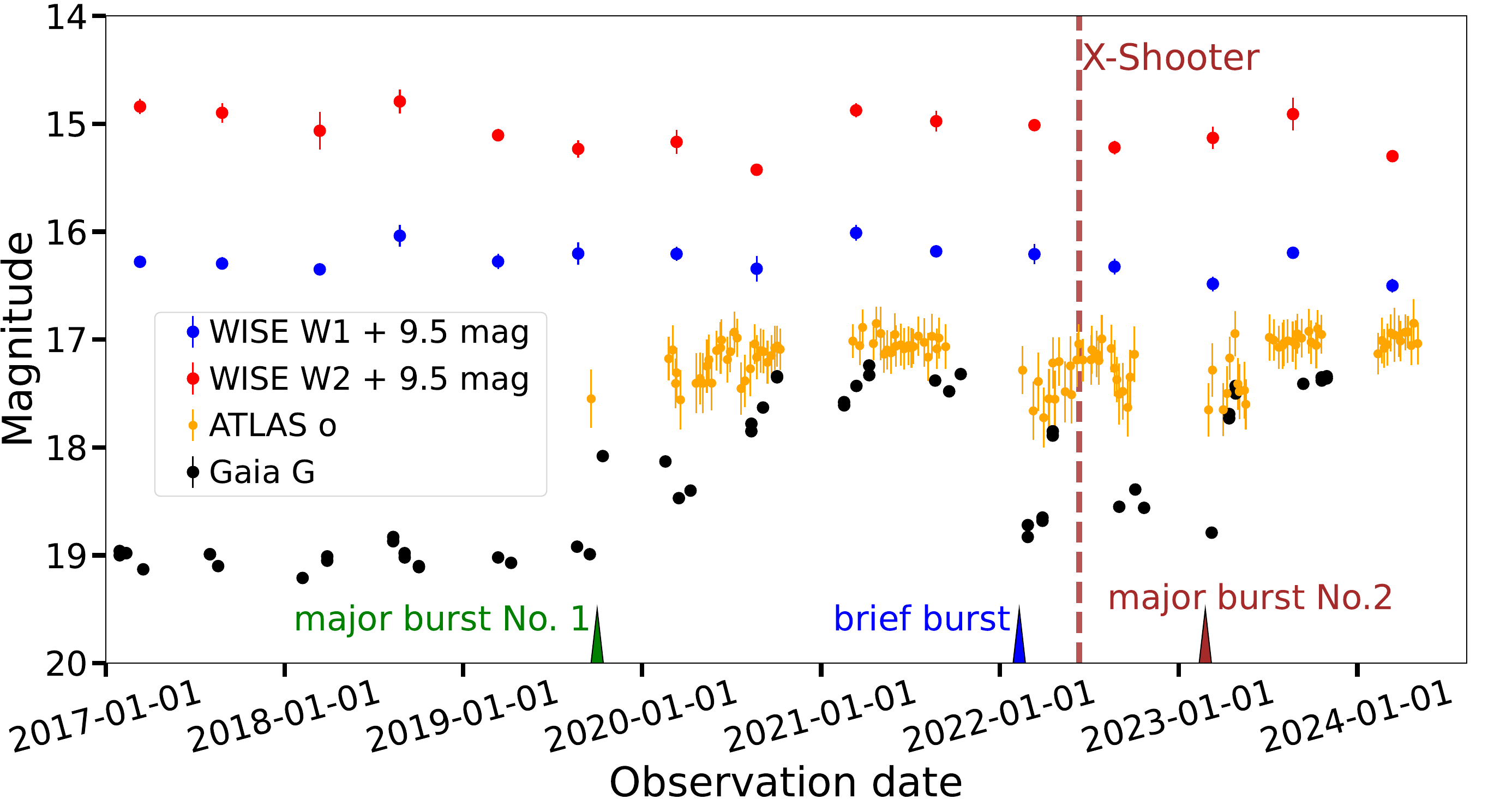}
    \caption{Gaia20dsk light curve. Gaia $G$ photometry is represented by black dots, with error bars similar in size to or smaller than the data points. ATLAS $o$ weekly median photometry is shown in orange. The WISE $W1$ and $W2$ bands are represented by blue and red dots, shifted by 9.5\,mag for clarity. The dashed vertical line indicates the date of the X-SHOOTER measurements. The three triangles at the bottom of the plot mark the first major burst, which began in 2019 October (green); a short brightening event around 2022 February (blue). A second major burst that started in 2023 February (brown).}\label{fig:light}
\end{figure}

The multiband light curve of the Gaia20dsk is shown in Fig.~\ref{fig:light}.
The quiescent magnitude of the star is $G\sim$19\,mag.
Gaia20dsk began brightening in 2019 October and continued brightening for approximately one-and-a-half years before reaching a peak brightness level of 17.2\,mag ($\Delta G$ = 1.8\,mag). 
Gaia20dsk remained at that level until late 2021, when it began to dim.
In 2022, the Gaia photometry showed a few bright points, suggesting another brightening event.
Starting in 2023 February, the source brightened for the third time at a much faster rate than during the previous episodes and reached a peak of 17.4\,mag.
For the Gaia measurements, we calculated the uncertainties using the empirical method first presented in \citet{Gaia_err}.
\setlength{\parskip}{0pt}

We complemented our Gaia light curve with data from the Asteroid Terrestrial-impact Last Alert System (ATLAS; \citealt{ATLAS}), utilizing reduced photometry from the ATLAS Forced Photometry server \citep{forced_photmetry}.
ATLAS has data in two photometric bands: $o$ ($\lambda_{\rm eff} = 662.98$ nm) and $c$ ($\lambda_{\rm eff} = 518.24$ nm).
Gaia20dsk was only detected in the $o$ band.
To ensure that we used only high-quality photometry, we applied selection criteria based on the $chi/N$ value (as reported by the ATLAS pipeline, which reflects the reduced $\rm  \chi^2$ from the point spread function (PSF), fitting of individual measurements), signal-to-noise ratio (S/N), and the width at half maximum (FWHM) of the PSF fit.
Details of this filtering process are provided in Appendix~\ref{app:filtering}.
After refining the dataset, we computed weekly median values to improve the clarity of the light curve.
The brightening in the $o$ band was confirmed after a careful visual inspection of the target and the surrounding stars.
The final $o$ band light curve is shown in Fig.~\ref{fig:light}.
When Gaia20dsk was in quiescence, it was not detected in the $o$ band. 
A brief brightening event that started in 2022 February (blue triangle in Fig.~\ref{fig:light}) was observed by ATLAS in the $o$ band, although it was not fully captured by Gaia.
The most recent brightening event, which began in early 2023, has remained in a plateau phase since mid-2023.

We constructed the mid-IR (MIR) light curve of Gaia20dsk using NEOWISE \citep{NEOWISE} data.
Given the relatively large beam size of WISE ($FWHM_{\rm W1} = 6.1\arcsec$, $FWHM_{\rm W2} = 6.4\arcsec$), flux contamination from foreground or background sources, or source blending, could affect the measured brightness \citep{Ribas2014,Dennihy2020}.
To assess the potential contamination, we examined higher-resolution images from 2MASS \citep{Skrutskie}, Spitzer \citep{Werner}, and VIRAC \citep{virac}.
We overplotted circles with radii of $1.3 \times FWHM_{W1}$ and $1.3 \times FWHM_{W2}$ around the coordinates of Gaia20dsk to simulate the WISE beam size (see Fig.~\ref{fig:wise}).
We defined this search area according to the WISE documentation\footnote{\url{https://wise2.ipac.caltech.edu/docs/release/allsky/expsup/sec4_4c.html}}.
The images revealed a stronger source near our defined search radius; however, the source is not entirely contained within it.
Additionally, none of the images underwent an active deblending process, as indicated by the WISE \texttt{Active deblending} quality parameter\footnote{\url{https://wise2.ipac.caltech.edu/docs/release/neowise/expsup/sec2_1a.html\#w1sat}}.  
We find that the influence of this source on our data is negligible, and thus no correction was applied (see Appendix~\ref{app:filtering2}).

After checking the images, we applied filtering based on the quality parameters provided by the NEOWISE-R Single Exposure (L1b) Source Table.  
The details of this filtering process are provided in Appendix~\ref{app:filtering2}.
Between 2020 August and 2021 March, Gaia20dsk brightened by 0.33~\,mag in $W1$ and by 0.5~\,mag in $W2$ (see Fig.~\ref{fig:light}). 
This event appears to have begun approximately five months after the first $G$ brightening. 
The brightening event after 2023 was not as clear in the WISE bands as it was in the optical bands. 
The variability of the WISE light curve is discussed in Sect.~\ref{sec:discussion}.

\section{Spectroscopic observations}\label{sec:data_reduction}
The spectrum of Gaia20dsk was obtained on 2022 June 12 using the X-SHOOTER instrument \citep{Vernet}, an echelle spectrograph mounted on the ESO Very Large Telescope at the Paranal Observatory in Chile (Program ID: 108.22LN; PI: Cruz-Sáenz de Miera).
By covering the spectral range from 300 to 2500 nm, this instrument can simultaneously observe the ultraviolet (UVB), visible (VIS), and NIR spectra.
We used the 0.5\arcsec$\times$11\arcsec, 0.4\arcsec$\times$11\arcsec, and 0.4\arcsec$\times$11\arcsec size slits, providing spectral resolutions of 9700, 18400, and 11600 in the UVB, VIS, and NIR arms, respectively.
The exposure times for UVB, VIS, and NIR regions were 1492\,s, 1536\,s, and 1600\,s, respectively.
Gaia20dsk was also measured with the 5\arcsec$\times$11\arcsec slit to correct for slit-losses and estimate the actual flux level by scaling the narrow slit spectrum to match the broad slit spectrum. 

\begin{figure*}[!htb]
    \includegraphics[width=\textwidth]{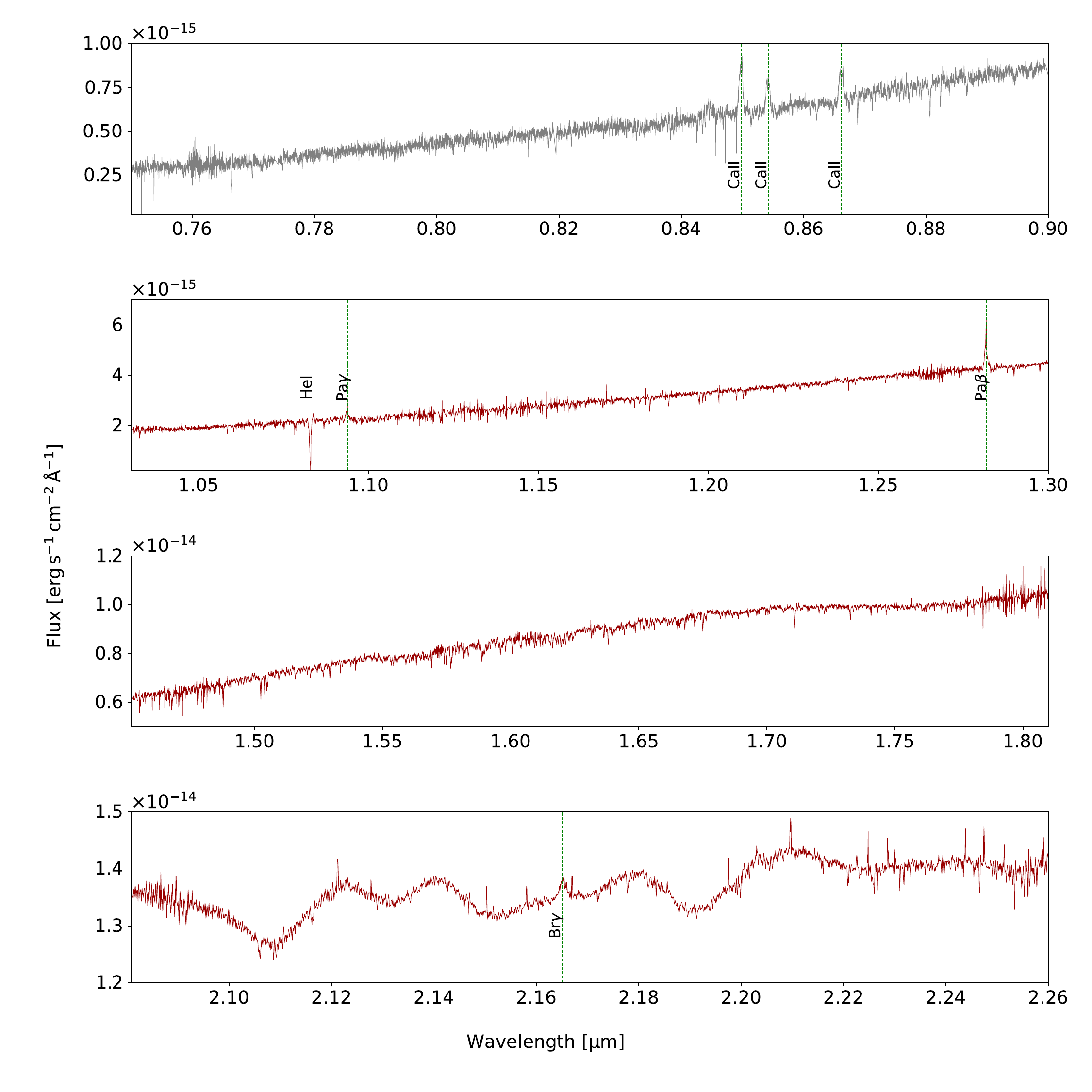}
        \caption{X-SHOOTER spectrum of Gaia20dsk. The optical part is shown in gray, while the IR range is shown in red. The spectrum is not corrected for extinction. The noisy part below 7500\,Å is not shown. \label{fig:spec}}
\end{figure*}

The data were reduced using the X-SHOOTER pipeline \citep{xshooter_pipeline}, and corrected for tellurics using MOLECFIT \citep{MOLECFIT1,MOLECFIT2}.
The median S/N in the UVB below 556\,nm is approximately 0.15, whereas in the VIS range below 750\,nm the median S/N is around 2.5. 
The spectrum (see Fig.~\ref{fig:spec}) shows an increasing flux density toward longer wavelengths and exhibits several emission lines, nine of which are known magnetospheric accretion tracers \citep{Alcala17}: H$\beta$, H$\alpha$, the \ion{Ca}{II} IR triplet, Pa$\delta$, Pa$\gamma$, Pa$\beta$, and Br$\gamma$.
H$\beta$ and H$\alpha$ are in the portion of the spectrum with low S/N; therefore, we cannot accurately estimate their line fluxes and caution should be exercised when interpreting them.

\subsection{Photosphere of Gaia20dsk: Identification and subtraction}\label{par:remove}
The shape and flux density of emission lines utilized to determine the mass accretion rate are contaminated by the stellar photosphere.
The spectral lines flux density and shape are influenced by the continuum level of the star, mostly due to veiling \citep{Hartmann16}.
To eliminate these effects, we subtracted the photospheric contribution from the Gaia20dsk spectrum.

To characterize the photosphere of Gaia20dsk, we compared its spectrum to stellar templates from the third data release of the X-SHOOTER Spectral Library (XSL; \citealt{Verro}), which contains 830 spectra of 683 stars. 
Because Gaia20dsk is likely a low- to intermediate-mass YSO, we selected template stars with properties characteristic of T~Tauri stars \citep{ttauri_star_protperties} and Herbig~Ae/Be stars \citep{Alecian2013, Montesinos2009}.
Specifically, we kept stars with effective temperatures between 4000 and 15000 K, surface gravities ($\log g$) between 3.0 and 4.5, and metallicities between $-0.5$ and $+0.5$. 
Applying these filters resulted in a subset of 125 XSL templates.

\begin{figure}[!htb]
  \centering
  \includegraphics[width=\linewidth]{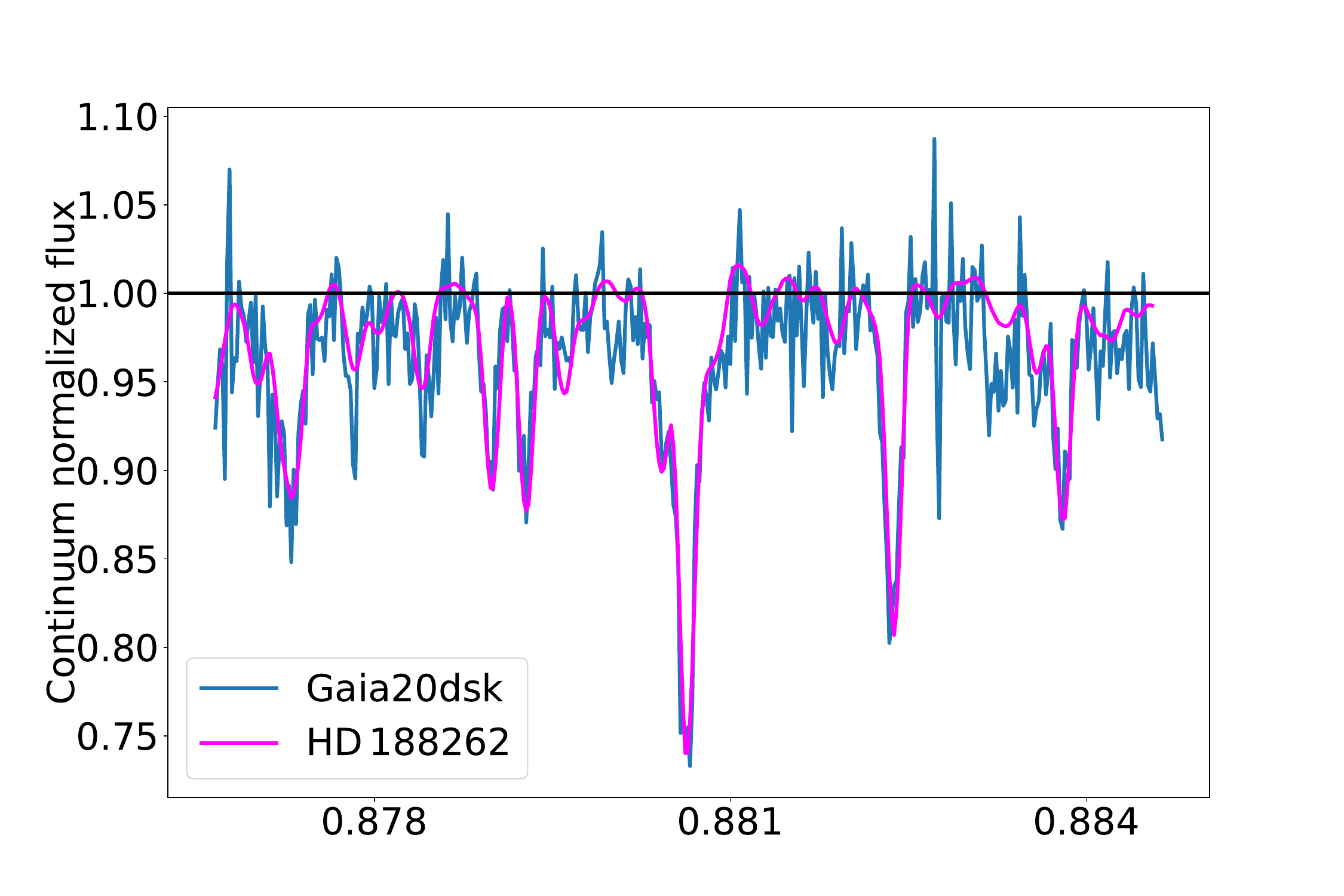}
  \includegraphics[width=\linewidth]{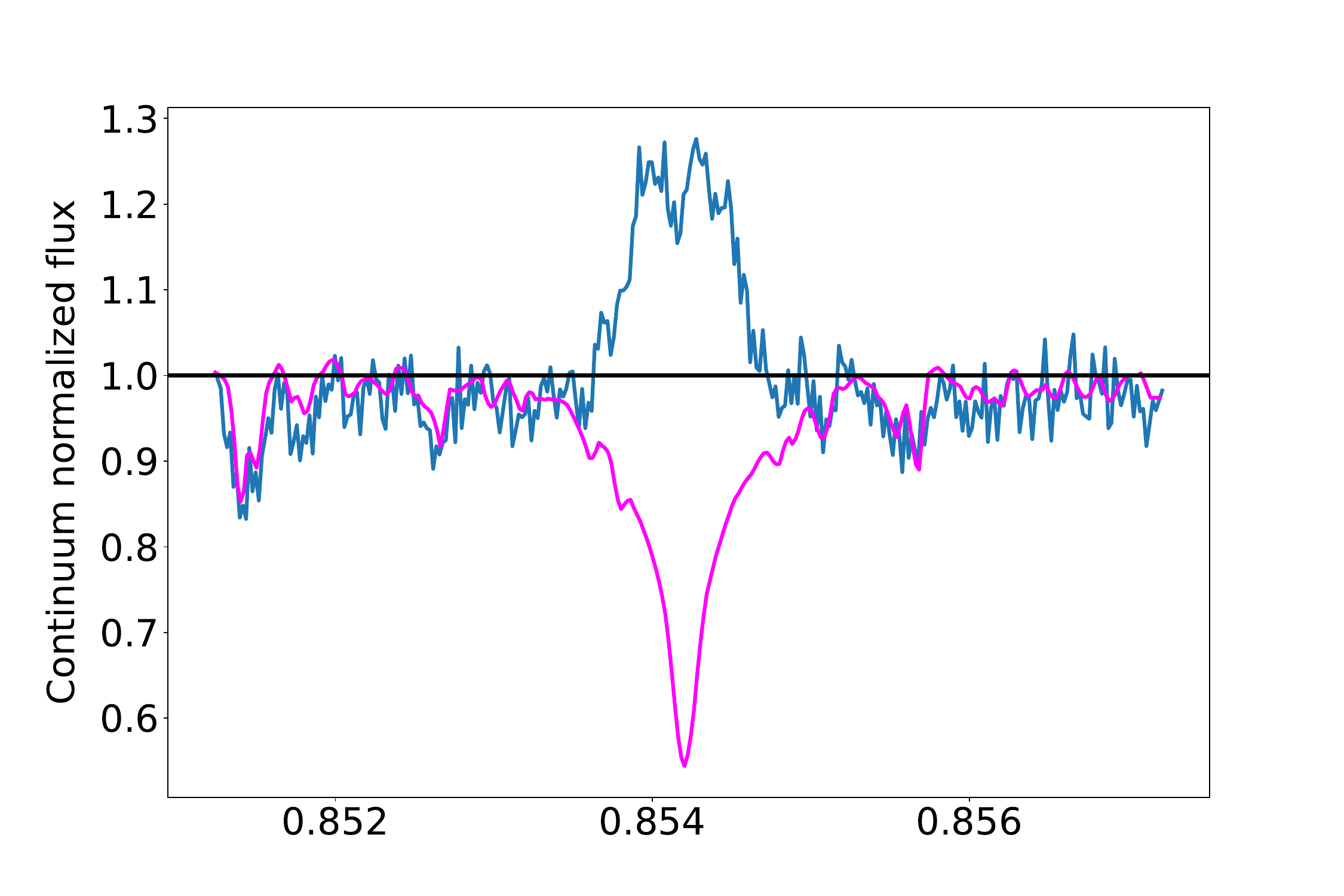}
  \includegraphics[width=\linewidth]{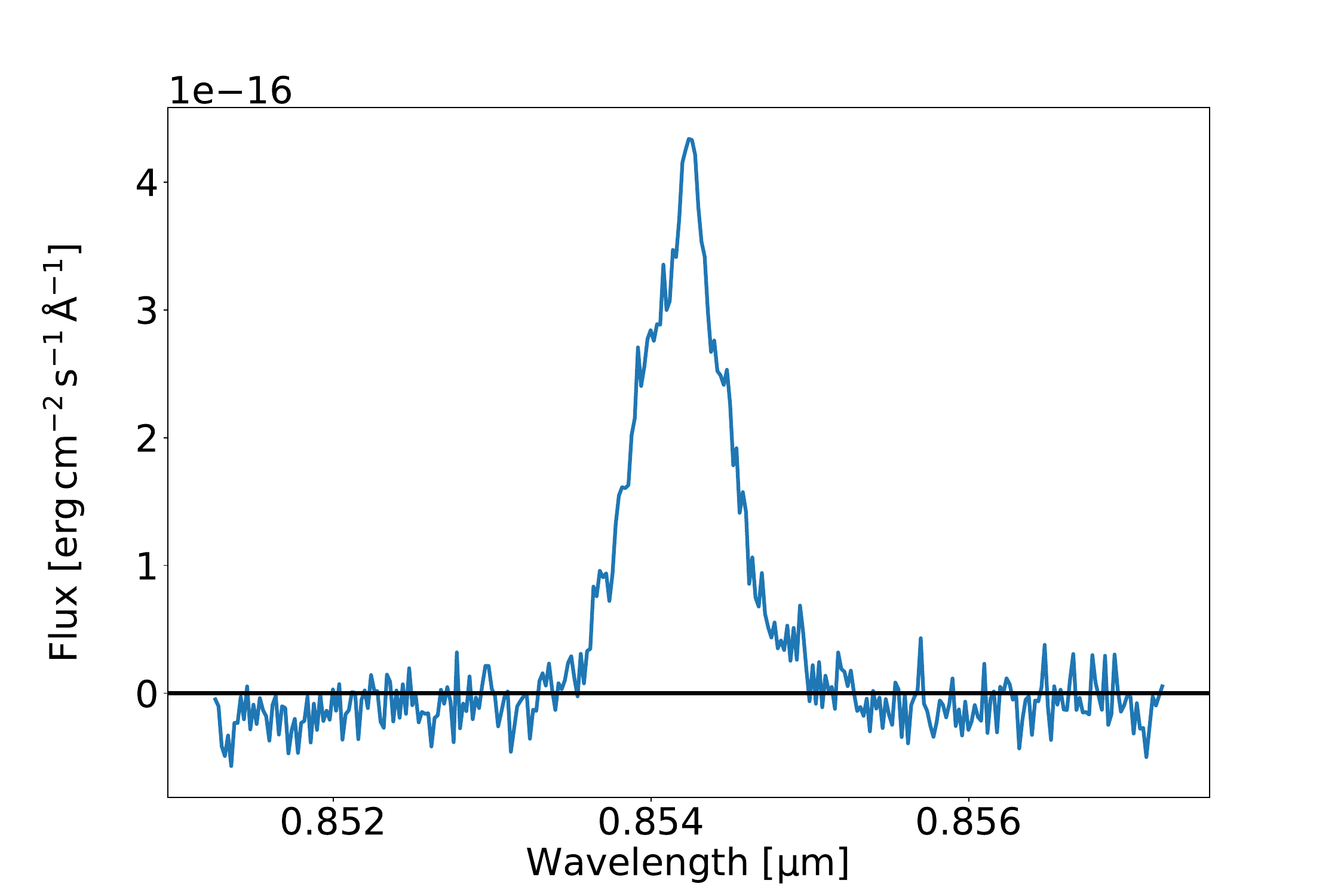} 
  \caption{Top panel: Segment of the spectra of our target Gaia20dsk (in blue) and the best-fitting photosphere HD188262 (shown in magenta). See Sect.~\ref{par:remove} for details. 
  Middle panel: Example of one of the \ion{Ca}{II} lines before correction with the best-fitting photosphere.
  The colors indicate the same sources as in the top panel.
  Bottom panel: Photosphere-subtracted accretion-tracing line.
  \label{fig:compare}}
\end{figure}

We used a portion of the optical spectrum (8770-8840 \AA) of Gaia20dsk to identify its photosphere. 
This section was chosen because it was the first region containing purely absorption lines (see Fig.~\ref{fig:compare}) and because the effects of veiling are less pronounced at these short wavelengths. 
The comparison between Gaia20dsk and the 125 templates from the XSL was performed using continuum-normalized spectra.
We applied the \textsc{fit\_continuum} function from the \textsc{specutils}\footnote{\url{https://specutils.readthedocs.io/en/stable/}} package to the 126 spectra (125 template + Gaia20dsk) in order to determine the continuum level for normalization.

Before finding the best match between Gaia20dsk and the 125 XSL stars, we accounted for the veiling and rotation of accreting stars by applying these effects to each template spectrum.
First, we applied a barycentric velocity correction (with the \textsc{astropy} Python package) for the spectral range of Gaia20dsk.
The template spectra were then shifted into the same velocity frame as Gaia20dsk, removing relative radial-velocity offsets and enabling a direct comparison of their spectral features.
The templates were subsequently broadened by increasing $v \sin i$ using the \texttt{rotBroad} function from the \texttt{PyAstronomy} package, which implements rotational broadening following \citet{Gray2005}.
Broadening was applied over grids of \( v \) and \( i \), with \( v \) sampled from 0 to 85\,km\,s\(^{-1} \) in steps of 15\,km\,s\(^{-1} \), and \( i \) from 0\(^\circ\) to 90\(^\circ\) in steps of 15\(^\circ\).
For each broadening step, we applied a series of veiling factors from 0 to 5 in steps of 0.1, using the following equation,
\begin{equation}\label{eq:veil}
    F_{\mathrm{temp}}^{\mathrm{veil}} = \dfrac{F_{\mathrm{temp}} + v_\lambda}{1 + v_\lambda},
\end{equation}
where $F_{\rm temp}$ is the continuum-normalized spectrum of the template, $v_\lambda$ is the veiling factor, and $F_{\rm temp}^{\rm veil}$ is the veiled flux density of the continuum-normalized template.

We compared the Gaia20dsk spectrum to each veiled and broadened template, and computed a $\chi^2$. 
Based on the $\chi^2$ minimization, two stars emerged as the best fitting templates with the same $\chi^2$: HD~116544, and HD~188262.
As a final decision, we adopted HD~188262. Using its effective temperature ($T_{\rm eff}$) together with the derived stellar luminosity of Gaia20dsk yields an age of 0.5–1.4 Myr (see Sect.~\ref{sec:anal}), consistent with the expected evolutionary stage of a Class I or flat spectrum (FS) source \citep{Evens2009}.
By contrast, adopting the $T_{\rm eff}$ of HD~116544 implies a much younger age (<0.2 Myr), characteristic of a Class 0 object \citep{Evens2009}.
Because Gaia20dsk is classified as a Class I or FS source (see Sect.~\ref{sec:szintetikus}) and its spectrum clearly shows photospheric absorption lines (which are not expected for Class 0 objects). Therefore, we adopted HD~188262 as the preferred template.

The parameters that minimize $\chi^2$ for HD~188262 are a veiling factor of $v_\lambda = 0.1$ and broadening factor of $v \sin i = 31\rm \,km\,s^{-1}$.
HD~188262 has an effective temperature ($T_{\rm eff}$) $5919\pm93$ K, with $\log{g}$ being 3.25, while [Fe/H] is 0.27 \citep{xslparam}.
We utilized this $T_{\rm eff}$ to determine the stellar parameters in Sect.~\ref{sec:parameter}.
In Fig.~\ref{fig:compare}, we show a portion of the Gaia20dsk spectrum alongside the best-fit template (top panel), the \ion{Ca}{II} 8542\,\AA{} line before subtraction (middle panel), and the same line after removal of the photospheric component (bottom panel).

\subsection{Line profiles}\label{sec:lines}
The accretion tracer lines include H$\beta$ ($\lambda_{\rm rest}$ = 0.486 \textmu m), H$\alpha$ ($\lambda_{\rm rest}$ = 0.656 \textmu m), the \ion{Ca}{II} triplet ($\lambda_{\rm rest}$ = 0.849 \textmu m, 0.852 \textmu m, 0.866 \textmu m), Pa$\delta$ ($\lambda_{\rm rest}$ = 1.005 \textmu m), Pa$\gamma$ ($\lambda_{\rm rest}$ = 1.094 \textmu m), Pa$\beta$ ($\lambda_{\rm rest}$ = 1.282 \textmu m), and Br$\gamma$ ($\lambda_{\rm rest}$ = 2.166 \textmu m). 
Because veiling is wavelength-dependent, we determined it individually for each spectral line before the photosphere subtraction.
For this, we used the photospheric template HD~188262 and the method described in Sect.~\ref{par:remove}.
We adopted the $v \sin i$ value derived in Sect.~\ref{par:remove} and we applied Eq.~\ref{eq:veil} to compute the veiling.
In practice, we measured the veiling from the nearby absorption features in the vicinity of each line, while excluding the emission components.
On average, the veiling at the end of the VIS range was found to be around $\nu_\lambda \sim 0.5$ ($\nu_{\mathrm{CaII}} = 0.5\pm0.1$), increasing toward the NIR, reaching $\nu_\lambda \sim 3.5$ ($\nu_{\mathrm{Pa\delta}} = 0.5\pm0.1$, $\nu_{\mathrm{Pa\gamma}} = 2.5\pm0.1$, $\nu_{\mathrm{Pa\beta}} = 3.5\pm0.6$) in the $J$ band and up to $\nu_\lambda \sim 10$ ($\nu_{\mathrm{Br\gamma}} = 10.1\pm2.5$) in the $K$ band.

After photosphere subtraction, most accretion tracer lines show slight blueshift of approximately $-10$ to $-15$ km s$^{-1}$, whereas \ion{Ca}{II} triplet lines show slight redshift of about $+15$ km s$^{-1}$.
These shifts are smaller than the velocity resolution of the spectra ($\Delta v_{\rm VIS} = 16$ km s$^{-1}$, $\Delta v_{\rm NIR} = 26$ km s$^{-1}$) and should therefore be interpreted with caution.
In the vicinity of Br$\gamma$ at 2.167~\textmu{}m and H$\beta$ at 0.486~\textmu{}m, we identified a spectral peak that could not be associated with any well-known line. Therefore, we treat it as an artifact and excluded it from further analysis.

Each of the corrected lines exhibits a single-peaked structure (see Figs. in Appendix~\ref{app:gaussians}).
To determine the amplitude, central wavelength, and width of the line profile, we fit Gaussian functions to each of the lines.
We fitted a single Gaussian for all lines using the Python \textsc{curve\_fit} function, which performs a nonlinear least-squares optimization.
In the cases where the line exhibited both broad and narrow components (\ion{Ca}{II} triplet, Pa$\delta$, Pa$\gamma$, Pa$\beta$, Br$\gamma$), we fit two Gaussian components (examples shown in Figs.~\ref{app:caII1}\,-\,\ref{app:brgamma}).
The best-fit values are shown in Table~\ref{tab:fwhm}.
\begin{table*}[!htb]
\centering
\caption{Best-fit parameters of the identified accretion-tracing lines.\label{tab:fwhm}}
\begin{tabular}{ccccc}
\hline
Line & Rest wavelength\tablefootmark{a}(\AA{}) & Peak position\tablefootmark{b} (\AA{}) & Velocity shift (km\,s$^{-1}$) & FWHM\tablefootmark{c}(km\,s$^{-1}$)\\
\hline
H$\beta$ & 4861.3 & 4861.1 & $-$17$\pm$2 & 60$\pm$2\\
H$\alpha$ & 6562.8 & 6562.6 & $-$9$\pm$1 & 47$\pm$1\\
\ion{Ca}{II} & 8498.0 & 8497.9 & $+$16$\pm$2 & [231$\pm$4, 82$\pm$5]\\
\ion{Ca}{II} & 8542.1 & 8541.9 & $+$16$\pm$1 & [247$\pm$2, 42$\pm$3]\\
\ion{Ca}{II} & 8662.1 & 8662.0 & $+$13$\pm$1 & [253$\pm$6, 40$\pm$4]\\
Pa$\delta$ & 10049.8 & 10049.04 & $-$13$\pm$1 & [209$\pm$8, 48$\pm$2]\\
Pa$\gamma$ & 10938.1 & 10937.8 & $-$7$\pm$1 & [245$\pm$9, 32$\pm$2]\\
Pa$\beta$ & 12818.1 & 12817.7 & $-$10$\pm$1 & [274$\pm$5, 44$\pm$1]\\
Br$\gamma$ & 21655.3 & 21653.4 & $-$8$\pm$1 & [200$\pm$4, 20$\pm$3]\\
\hline
\end{tabular}
\tablefoot{
\tablefoottext{a}{The wavelengths are measured in air.}
\tablefoottext{b}{Where the line shows both double and narrow components, we report the peak position for the narrow component.}
\tablefoottext{c}{When the line exhibits both broad and narrow components, we report each FWHM separately. In such cases, the column contains two values within brackets, where the first value corresponds to the broad component and the second to the narrow component.}}
\end{table*}

The combined presence of a broad and narrow component is common in CTTS.
The narrow component is thought to originate in the stellar chromosphere, while the broad component is associated with the infalling material \citep{Herbig}. 
For Pa$\delta$, we smoothed the spectrum to improve the S/N, as it was noisier than the other lines, complicating the continuum determination.
The smoothing was performed by taking the mean flux density over three pixel wide bins.

The two Balmer lines detected in our spectra, H$\alpha$ and H$\beta$, both exhibit notably narrow profiles compared to the other emission lines ($\sim$250 km\,s$^{-1}$ on average).
H$\beta$ has a FWHM of 60$\pm$1 km\,s$^{-1}$ (Fig.~\ref{app:hbeta}), whereas H$\alpha$ is even narrower, with a FWHM of 47$\pm$1 km \,s$^{-1}$ (Fig.~\ref{app:halfa}).
The red wing of H$\alpha$ displays shallow, elongated, and very weak emission features, which might be attributed to outflowing winds.

Among the accretion tracers, the \ion{Ca}{II} triplet lines are the strongest emission lines in the NIR spectra (Figs.~\ref{app:caII1}\,-\,\ref{app:caII3}). 
Furthermore, the \ion{Ca}{II} 8498\,\text{\AA{}} line has the narrowest FWHM (231$\pm$4 km\,s$^{-1}$) among the triplet, while \ion{Ca}{II} 8542\,\text{\AA{}} and \ion{Ca}{II} 8662\,\text{\AA{}} display similar FWHMs (247$\pm$2 km\,s$^{-1}$ and 253$\pm$6 km\,s$^{-1}$).
Additionally, \ion{Ca}{II} 8498 \text{\AA} is the strongest among the triplet, while \ion{Ca}{II} 8662 is the weakest.
A similar tendency was previously noted, for example by \citet{HartmannFred1992} in their analysis of the \ion{Ca}{II} triplet.

In addition to the previous accretion-tracing lines, the spectrum also exhibits several jet-tracing lines, such as [\ion{S}{II}]~6716\,\AA{} and [\ion{S}{II}]~6731\,\AA{}, as well as outflow indicators such as [\ion{N}{II}]~6548\,\AA{} and [\ion{N}{II}]~6583\,\AA{} (see Figs.~\ref{fig:nii_2_line}\,-\,\ref{fig:sii_2_line}).
We note that a double-peaked structure is visible in some forbidden lines, similar to that observed in EX Lupi during its outburst \citep[e.g.,][]{2010ApJ...719L..50A}.
We detected the \ion{He}{I}~10\,830\,\text{\AA{}} (Fig.~\ref{fig:hei_line}), which exhibits a broad absorption component that might arise from a stellar wind \citep[e.g.,][]{Erkal} or from an outflow \citep[e.g.,][]{Kwan2011}.
The spectrum also shows the H${\rm _2}$ 21218 \text{\AA{}} emission line (see Fig.~\ref{fig:h2_line}), which is associated with shocks from molecular outflows \citep{MNOR2017}.
We consider these lines to be beyond of the scope of this paper and are not discussed further.

\section{Spectral energy distribution}\label{sec:szintetikus}
We compiled the SED for Gaia20dsk using archival data from the VO SED Analyzer (VOSA) services \citep{Bayo} and VizieR \citep{vizier}.
It includes archival photometric data from various surveys, including Gaia \citep{gaia}, 2MASS \citep{Skrutskie}, WISE \citep{Wright}, Spitzer \citep{Werner}, DENIS \citep{denis}, VIRAC \citep{virac}, and VPHAS+ \citep{vphaps+}. 
The flux densities used for the SED are listed in Appendix~\ref{sec:sed_flux}.

Figure~\ref{fig:SED} shows the full SED, with different instruments indicated by distinct colors.
The X-SHOOTER spectrum is slightly smoothed for clarity, and synthetic photometry was calculated in the $I$, $J$, $H$, and $K_{\rm s}$ bands to compare with the archival photometric data.
\begin{figure}[!htb]
    \centering
    \includegraphics[width=\linewidth]{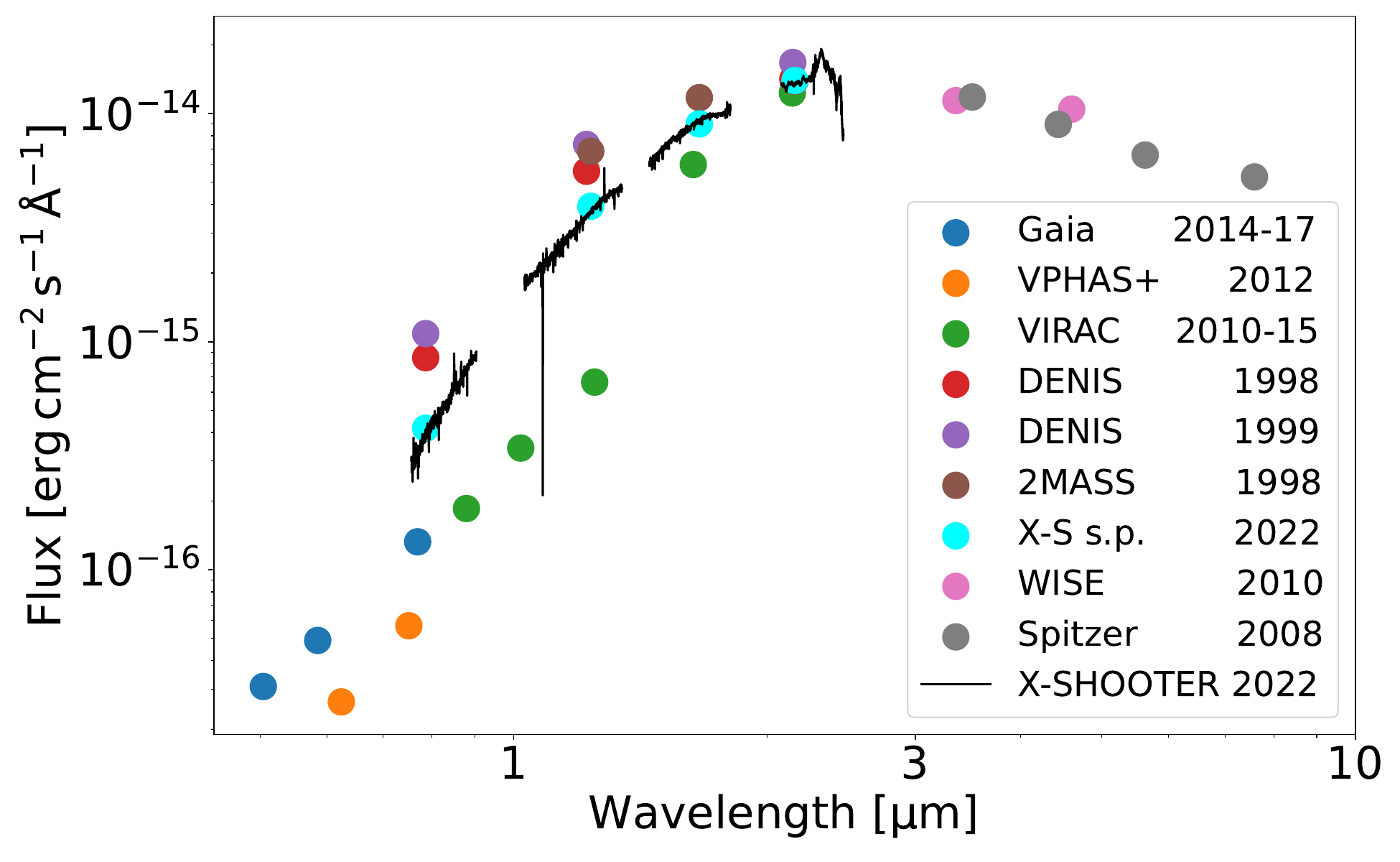}
    \caption{Gaia20dsk SED. Different colored dots denote different surveys, as explained in the legend. The synthetic photometry in $I$, $J$, $H$, and $K_{\rm s}$ bands represented by the cyan color were computed from the X-SHOOTER spectrum (labeled X-S s.p.). The spectrum itself is plotted with reduced resolution for clarity. The points are not corrected for extinction. The sizes of the error bars are comparable to or smaller than the symbol size. \label{fig:SED}}
\end{figure}

Gaia20dsk shows differences exceeding 3$\sigma$ between synthetic photometry and archival data in the DENIS $I$ and $J$ bands and the 2MASS $J$ and $H$ bands. 
In the DENIS $J$ and 2MASS $J$ bands, the source dimmed by $0.24\pm0.1$\,mag between 1998 August 10 and October 29, then brightened by $0.29\pm0.1$\,mag by 1999 July 22. 
Without supporting spectroscopy, the significance of these events and whether they are due to changes in accretion or extinction remain unclear.

We applied several photometric methods to classify Gaia20dsk. 
Using 2MASS and WISE measurements, we computed $[H]-[K_s]=1.24\pm0.03$\,mag and $[W1]-[W2]=1.32\pm0.07$\,mag, suggesting a Class I FS object when compared to Fig.~6 of \citet{Koenig14}.
Spitzer colors $[I1]-[I2]=0.685\pm0.07$, $[I1]-[I3]=1.357\pm0.04$, $[I2]-[I3]=0.672\pm0.07$, and $[I2]-[I4]=1.74\pm0.07$\,mag meet the criteria defined by \citet{Gutermuth2008} for a Class II star or slightly earlier. 
\citet{spectral_index} provided three equations for determining the spectral index $\alpha$ \citep{Lada} using NIR photometry between 4 and 24~\textmu{}m, where interstellar extinction is reduced.
Although we lack measurements near 24~\textmu{}m, we used the Spitzer $[I2]-[I4]$ color and Equation 9 of \citet{spectral_index}:
$\alpha_{[I2]-[I4]}\approx1.64([I2]-[I4])-2.82$.
This gives a slope of $\alpha_{[I2]-[I4]} = 0.04 \pm 0.07$, consistent with a FS classification ($-0.3 \le \alpha < 0.3$).
Considering all three methods, Gaia20dsk has been classified as Class I or FS source.

\begin{figure}[!htb]
    \centering
    \includegraphics[width=1\columnwidth]{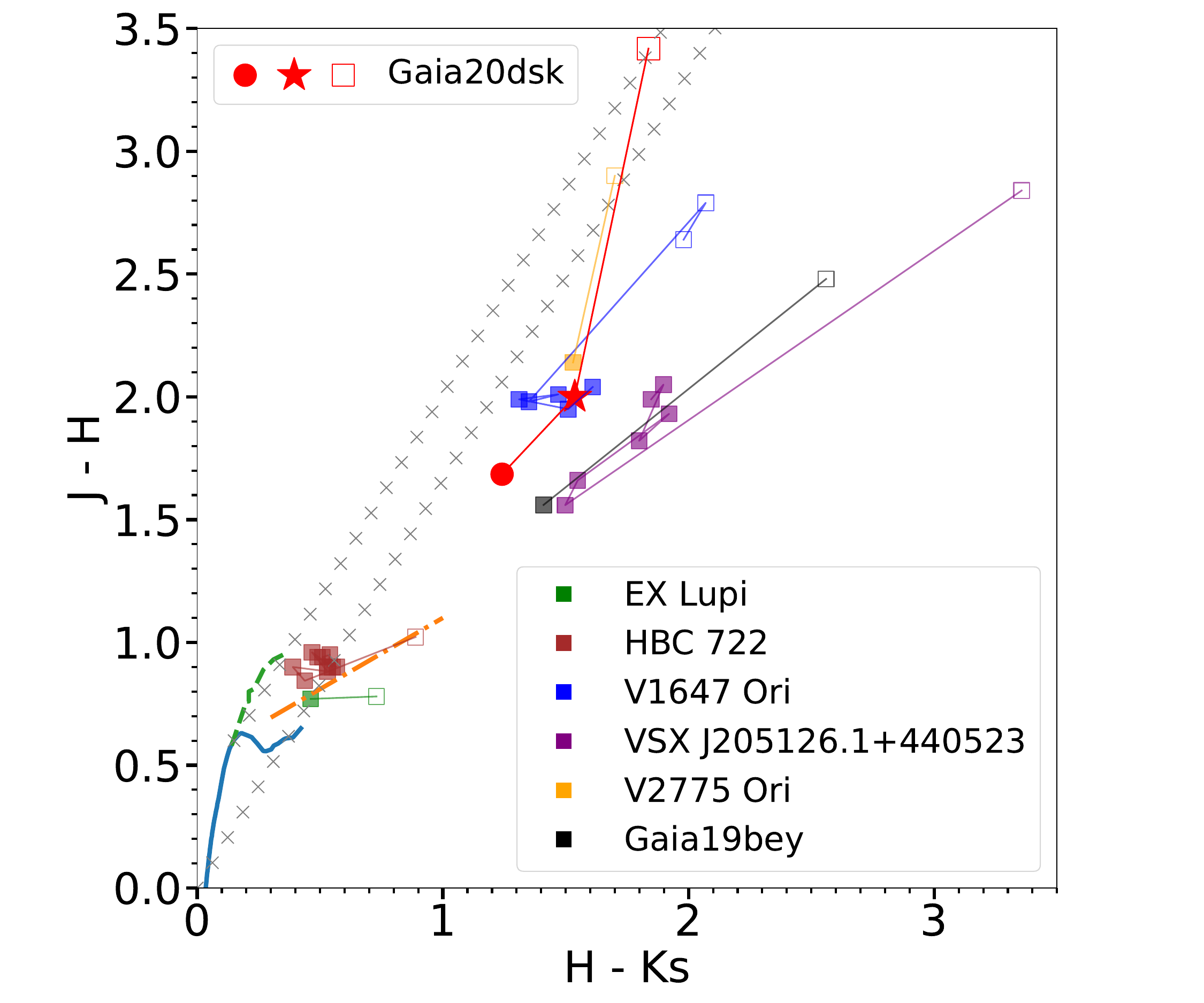}
    \caption{$J-H$ versus $H-K_{\rm s}$ diagram. 
    The main sequence is represented by a blue thick solid line \citep{Mamajek}, the giant branch by a green dashed line \citep{Koornneef1983}, and the reddening path (with $A_{\rm V} = 1$ mag steps) by gray cross symbols \citep{Cardelli}.
    The T Tauri locus is shown as an orange dash-dotted line \citep{Meyer1997}. Open squares denote quiescent colors, while filled squares represent outburst colors.
    For Gaia20dsk, the filled dot corresponds to the 2MASS measurements from 1998, the filled square to the VIRAC measurements (2010-2015), and the star symbol indicates synthetic photometry derived from the X-SHOOTER spectrum (2020). The error sizes are smaller than or comparable to the symbol sizes.}
    \label{fig:ccd_2mass}
\end{figure}

Objects occupy distinct regions in color--color diagrams and the position of a YSO in the $J,H,K$ space provides insight into its evolutionary stage \citep[see e.g.,][]{Lada1992}. 
Comparing this position with the extinction vector allows us to assess whether light variation could be caused by extinction.
In Fig.~\ref{fig:ccd_2mass}, the measured $J, H, K$ colors of Gaia20dsk are compared against those of other eruptive young stars: V1647~Ori \citep{Acosta}, EX~Lupi \citep{Juhasz2012}, HBC~722 and VSX~J205126.1+440523 \citep{Kospal11}, V2775~Ori \citep{Garatti2011}, and Gaia19bey \citep{Hodapp2020}.
Gaia20dsk 2MASS colors are depicted by a red filled circle, while the red star symbol represents the $JHK_{\rm s}$ synthetic photometry we calculated from the X-SHOOTER spectrum and the open red square represents the VIRAC measurements.
We note that to match the VIRAC $J$, $H$, $K_{\rm s}$ data with 2MASS, we applied a color correction for VIRAC bands using the formulas derived by \citet{2massvista}.
The cross symbols in the figure indicate the direction of extinction, signifying a measure of 1 magnitude of light attenuation in the $V$ band.
The red solid line connecting the 2MASS (filled red circle) and synthetic photometry (red star symbol) data points deviates from the extinction vector; thus, the discrepancy could be attributed to intrinsic variability.
In this case, we used $R_{\rm V} = 3.1$ total-to-selective extinction ratio, but using different values (e.g., $R_{\rm V} = 5.1$) does not affect this result.
Gaia20dsk shows a similar trend to other sources, as it becomes bluer during bursts, a behavior we discuss in detail in Sect.~\ref{sec:discussion}.

\begin{figure}[!htb]
    \centering
    \includegraphics[width=0.49\textwidth]{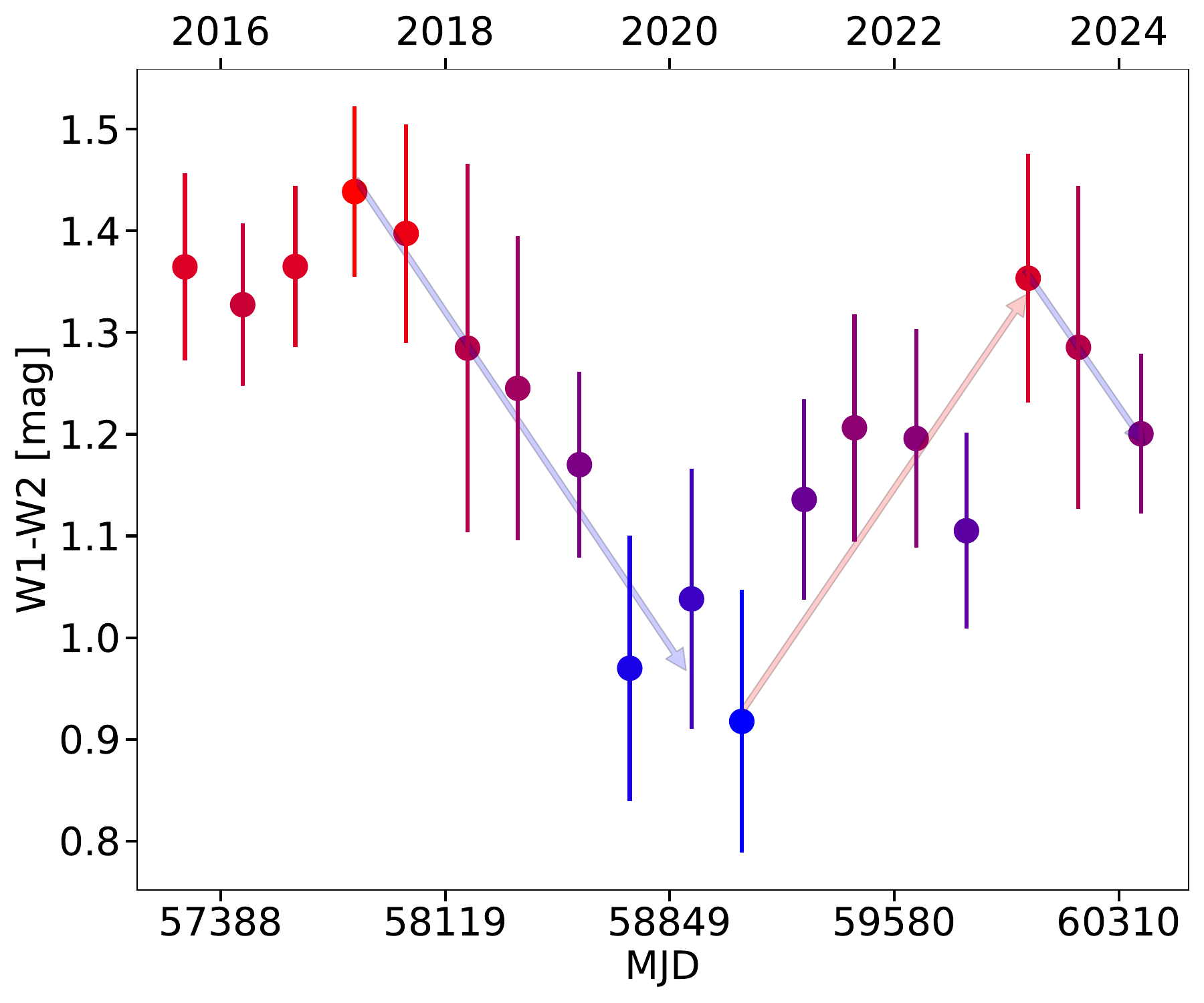}
    \caption{WISE $W1 - W2$ color curve. The panel illustrates how Gaia20dsk became bluer approaching the first major brightening event. Following this, two episodes of blueward color changes occur, superimposed on an overall reddening trend, which coincide with the brief and second major bursts.}
    \label{fig:wise_clc}
\end{figure}

In Fig.~\ref{fig:wise_clc}, we present the NEOWISE $W1-W2$ color curve, which shows Gaia20dsk becoming bluer leading up to the first major brightening event. 
The star reaches its maximum bluest color at the peak of this brightening event, which occurred between 2020 and 2021.
Afterward, the star appears to begin to follow a trend of becoming redder, although this  process was interrupted by two events during which the star became bluer. 
These bluer periods coincide with the brief burst and then the second major burst.

\begin{table*}[!h]
\caption{Calculated accretion properties based on the SED and derived stellar parameters.}
\centering
\begin{tabular}{l c c c c c @{\hspace{1.5cm}} c c c c}
         & $A_{\rm V}$ & $L_{\mathrm{bol,\,out}}^{\mathrm{SED}}$ & $L_{\mathrm{bol,\,qui}}^{\mathrm{SED}}$ & $L_{\mathrm{acc}}^{\mathrm{SED}}$ & $M_{\mathrm{acc}}^{\mathrm{SED}}$  
         & $\rm L_{\star}$ & $\rm R_{\star}$ & $\rm M_{\star}$ & age \\
         & [mag] & [$\rm log(L_\odot$)] & [log($\rm L_\odot$)] & [log($\rm L_\odot$)] & [log($\rm M_\odot\,\mathrm{yr}^{-1}$)] 
         & [$\rm L_\odot$] & [$\rm R_\odot$] & [$\rm M_\odot$] & [yr] \\
\hline
$d_{\rm NGC}$ & 4.3 & $2.00 \pm 0.01$ & $1.96 \pm 0.01$ & $0.90 \pm 0.10$ & $-6.2 \pm 0.2$ 
             & $45^{+15}_{-12}$ & $6^{+0.8}_{-0.7}$ & $3.2^{+0.1}_{-0.4}$ & $(1.4^{+0.6}_{-0.1}) \times 10^{6}$ \\

$d_{\rm BJ}$  & 5.4 & $2.41 \pm 0.23$ & $2.36 \pm 0.23$ & $1.41 \pm 0.90$ & $-5.6 \pm 0.9$ 
             & $110^{+88}_{-49}$ & $9.7^{+2}_{-4}$ & $4.3^{+0.7}_{-1.2}$ & $(5^{+8}_{-0.1}) \times 10^{5}$ \\
\hline
\end{tabular}
\label{tab:stellprop}
\end{table*}

\begin{table*}[]
\centering
\caption{Accretion-tracing lines used to determine $L_{\rm acc}$ and $M_{\rm acc}$.}
\begin{tabular}{cccccccc} 
 & & \multicolumn{3}{c}{$d_{\rm NGC}$} & \multicolumn{3}{c}{$d_{\rm BJ}$} \\
\multicolumn{2}{c}{} & \multicolumn{3}{c}{\rule{0.3\linewidth}{0.8pt}} & \multicolumn{3}{c}{\rule{0.3\linewidth}{0.8pt}} \\[3pt]
Line & $f_{\rm line}$ & $L_{\rm line} \times 10^{+24}$ & $L_{\rm acc}$ & $M_{\rm acc} \times 10^{-07}$ & $L_{\rm line} \times 10^{+25}$ & $L_{\rm acc}$ & $M_{\rm acc} \times 10^{-06}$ \\
 &  [W\,m$^{-2}$] & [W] & [$\rm L_{\odot}$] & [$\rm M_{\odot}$\,yr$^{-1}$] & [W] & [$\rm L_{\odot}$] & [$\rm M_{\odot}$\,yr$^{-1}$] \\ \hline
Pa$\delta$&             $3.8\times10^{-17}$ &           $1.2$&          $5.1^{+10.4}_{-3.4}$&           $4.1^{+8.5}_{-2.8}$ & $0.3$&                $15.0^{+29.3}_{-9.9}$&          $1.3^{+2.6}_{-0.9}$\\ 
Pa$\gamma$&             $6.0\times10^{-17}$ &           $1.9$&          $5.4^{+5.5}_{-2.7}$&            $4.4^{+4.5}_{-2.2}$ & $0.5$&                $16.0^{+15.4}_{-7.8}$&          $1.4^{+1.4}_{-0.7}$\\ 
Pa$\beta$&              $1.2\times10^{-16}$ &           $4.0$&          $4.5^{+6.0}_{-2.6}$&            $3.7^{+4.9}_{-2.1}$ & $0.9$&                $11.3^{+14.5}_{-6.4}$&          $1.0^{+1.3}_{-0.6}$\\ 
Br$\gamma$&             $2.0\times10^{-17}$ &           $0.7$&          $5.4^{+15.1}_{-4.0}$&           $4.4^{+12.3}_{-3.2}$ & $0.1$&                $11.3^{+30.6}_{-8.3}$&          $1.0^{+2.7}_{-0.7}$\\ 
\rule{0pt}{2.5ex}\\ 
 \hline\hline 
 \rule{0pt}{2.5ex} \\Pa$\delta$&                $3.9\times10^{-17}$ &           $1.3$&          $7.8^{+5.8}_{-3.3}$&            $6.2^{+4.6}_{-2.7}$ & $3.0$&                $23.1^{+15.6}_{-9.3}$&          $2.1^{+1.4}_{-0.8}$\\ 
Pa$\gamma$&             $6.0\times10^{-17}$ &           $2.0$&          $7.5^{+5.1}_{-3.0}$&            $5.9^{+4.1}_{-2.4}$ & $4.6$&                $22.1^{+13.6}_{-8.4}$&          $2.0^{+1.2}_{-0.8}$\\ 
Pa$\beta$&              $1.2\times10^{-16}$ &           $4.0$&          $9.3^{+5.1}_{-3.3}$&            $7.4^{+4.1}_{-2.6}$ & $9.4$&                $27.7^{+13.4}_{-9.0}$&          $2.5^{+1.2}_{-0.8}$\\ 
Br$\gamma$&             $1.6\times10^{-17}$ &           $0.5$&          $5.4^{+6.6}_{-3.0}$&            $4.3^{+5.2}_{-2.4}$ & $1.2$&                $16.6^{+18.2}_{-8.7}$&          $1.5^{+1.6}_{-0.8}$\\ 
\end{tabular}
 \label{tab:minden}
 \tablefoot{Left side of the table corresponds to calculations based on $d_{\rm NGC}$ and the right side to calculations based on $d_{\rm BJ}$. Data above the double line use  coefficients from \citet{Alcala14}, while data below the line use the coefficients from \citet{Fairlamb17}. See text for more details. The uncertainties for line fluxes and line luminosities are not listed as they are below 1\%.}\label{tab:mass}
 \end{table*}{} 
\section{Analysis}\label{sec:anal}
In this section, the stellar parameters of Gaia20dsk (e.g.,  luminosity, radius, age, and mass) as well as its accretion properties (e.g., accretion luminosity and accretion rate) are reported.
For outbursting young stars, photometric data suggest an accretion outburst, but do not provide a confirmation.
To confirm these outbursts, it is necessary to estimate the mass accretion rate during the bright phases of the event and compare it to an accretion rate measured during quiescence (if possible) or to a standard quiescent mass accretion rate.
To calculate the mass accretion rate ($M_{\rm acc}$), we first need to estimate the accretion luminosity ($L_{\rm acc}$).
We applied two methods to determine this:
one  based on the integration of the SED and the other on empirical relationships between the accretion luminosity and the luminosity of accretion tracer spectral lines.

\subsection{Accretion luminosities from the SED}\label{sec:laccSED}

As a first approach, we used the quiescence and the burst SEDs to estimate $L_{\rm acc}^{\rm SED}$.
Given that our source is classified as Class~I or FS, the bolometric luminosity in quiescence, $L_{\mathrm{bol,\,qui}}^{\mathrm{SED}}$, arises from multiple components: the stellar photosphere, the accretion luminosity (which persists even during quiescence), and the surrounding circumstellar material, including the disk and envelope.
During a burst, we assume the bolometric luminosity, $L_{\mathrm{bol,\,out}}^{\mathrm{SED}}$, can be expressed as the sum of this quiescent luminosity and an additional accretion luminosity component, $L_{\mathrm{acc}}^{\mathrm{SED}}$, resulting from enhanced accretion,
\begin{align}
L_{\mathrm{bol,\,out}}^{\mathrm{SED}} = L_{\mathrm{bol,\,qui}}^{\mathrm{SED}} + L_{\mathrm{acc}}^{\mathrm{SED}}.
\end{align}

The burst SED was compiled using synthetic photometry, while the quiescent SED was constructed from VIRAC measurements.  
In both cases, WISE and Spitzer data were used for wavelengths beyond the $K$ band, assuming that the SED shape remained unchanged at these longer wavelengths.
This assumption was supported by the fact that the $K$-band photometry remained largely unchanged between burst and quiescence.
Bolometric luminosities were estimated by fitting the data points with a spline function and integrating in the $\log \lambda$–$\log F_\lambda$ plane.
The longest available wavelength is 8 \textmu{}m from Spitzer.
Beyond this, we assumed the flux declines as $1/\lambda^2$ and we extrapolated the SED out to 100 \textmu{}m.
These bolometric luminosities should be considered lower limits because the lack of long-wavelength coverage may underestimate emission from possible colder material.
Before integrating, we corrected for foreground extinction using interstellar $A_V$ values from the DECaPS dust map \citep{Zucker2025}.
We did not correct for local extinction because that surrounding material re-radiates absorbed energy at longer wavelengths.
Afterward, we integrated the SED and computed the bolometric luminosity using $L_{\rm bol} = 4\pi d^2 F$,
where $d$ is the distance, and $F$ is the integrated flux.

Based on the calculations and assuming $d_{\rm NGC} = 1650^{+14}_{-7}\,\rm pc$, the bolometric luminosity increased from 92 to 100~L$_\odot$ as a result of the burst.  
Similarly, assuming $d_{\rm BJ} = 2542^{+704}_{-606}\,\rm pc$, it increased from 231 to 257~L$_\odot$.  
These changes correspond to accretion luminosities of 8 and 26~L$_\odot$, respectively.
The calculated bolometric and accretion luminosities, as well as the $A_{\rm V}$ values used, are listed in Table \ref{tab:stellprop}.

\subsection{Accretion luminosities from empirical relationships}\label{sec:lassALCALA}
To compare the \( L^{\rm SED}_{\rm acc} \) values computed in the previous subsection, we derived accretion luminosities under the assumption that mass is delivered onto the star via magnetospheric accretion.
We measured the line fluxes, \( f_{\rm line} \), of four accretion-tracing lines: Pa$\delta$, Pa$\gamma$, Pa$\beta$, and Br$\gamma$, using the \texttt{line\_flux} function from the \textsc{specutils} Python package.
In these calculations we did not use the H$\alpha$, H$\beta$ due to their potential contamination by non-accretion-related processes such as outflows. Additionally, we excluded the \ion{Ca}{ii} triplet lines (8498\AA, 8542\AA, 8662\AA) following the \cite{Alcala17} and \cite{Fairlamb17} recommendations, as these lines are less reliable for deriving $L_{acc}$.
The line luminosities were then calculated as $L_{\rm line} = 4\pi d^2 f_{\rm line},$
where \( d \) is the distance to the star and \( f_{\rm line} \) is the extinction-corrected flux.

The accretion luminosities were derived from \( L_{\rm line} \) using empirical relationships of the form 
\begin{equation}
\log\left(\frac{L_{\rm acc}}{L_\odot}\right) = a \log\left(\frac{L_{\rm line}}{L_\odot}\right) + b,
\end{equation}

where the coefficients \(a\) and \(b\) were adopted from \citet{Alcala17} and \citet{Fairlamb17}, who calibrated these relations for low-mass and intermediate-mass protostars, respectively.

Before applying these relations, we corrected the line fluxes not only for foreground extinction, but also for local extinction toward Gaia20dsk.
As no extinction value was previously reported for this source, we used the accretion luminosities to estimate \( A_V \) toward Gaia20dsk. 
In the absence of extinction (\( A_V = 0 \) mag), all accretion-tracing lines are expected to yield the same \( L_{\rm acc} \). 
Thus, we computed \( L_{\rm acc} \) without extinction correction and searched for the value of \( A_V \) for which the corrected line luminosities produced consistent accretion luminosities across the different tracers. 
Extinction corrections were performed using the \citet{Cardelli} reddening law with \( R_V = 3.1 \), implemented via the \textsc{CCM89} routine from the \textsc{dust\_extinction} Python package. 
Our estimated extinction is \( A_{\rm V} = 8.6 \pm 1.7 \) mag, which we used during the calculations. 
Since the distance acts as a uniform scaling factor applied equally to all line luminosities, the two distance estimates yielded the same extinction value. 
We found that the derived \( A_{\rm V} \) is consistent regardless of whether the \citet{Alcala17} or \citet{Fairlamb17} coefficients are used to compute \( L_{\rm acc} \).

In addition, we estimated the visual extinction by projecting the source position from the 2020 measurements in the $J-H$ vs.\ $H-K_{\rm s}$ color--color diagram (red star in Fig.~\ref{fig:ccd_2mass}) onto the locus of unreddened CTTS \citep{Meyer1997} along the reddening vector using the extinction law of \citet{Cardelli}. 
This gave $A_V = 8.9 \pm 1.9$\,mag, consistent with the values obtained from the empirical relations.
The measured line fluxes, line luminosities, and accretion luminosities for each tracer are listed in Table~\ref{tab:mass}.

\subsection{Stellar parameters and mass accretion rate}\label{sec:parameter}
To obtain accurate estimates of \( M_{\rm acc} \), we first need to constrain the stellar parameters, including stellar luminosity.
The stellar luminosities were estimated following the method of \citet{Fiorellino2021}, which uses the $K$ band photometry and bolometric correction to estimate the stellar luminosity. 
Because Gaia20dsk is a Class~I source, its observed flux contains contributions from both the star and the circumstellar disk and envelope. 
To account for this excess, we included the $K$ band veiling parameter, \( r_{\rm K} \), in the calculation.
This veiling was determined following the method described in Sect.~\ref{par:remove}, adopting the $v \sin i$ value derived in Sect.~\ref{par:remove} and using the spectral region around the Br$\gamma$ line (2.162--2.172~\textmu m), excluding the emission component itself.
The result is \( r_{\rm K} = 10.1 \pm 2.5 \).
We computed the bolometric magnitude as
\begin{align}
    M_{\rm bol} = BC_{\rm K} + m_{\rm K} + 2.5 \log(1 + r_{\rm K}) - A_{\rm K} - 5 \log\left(\frac{d}{10\,\mathrm{pc}}\right),
\end{align}
where \( BC_{\rm K} \) is the bolometric correction, 
\( m_{\rm K} \) is the brightness, 
\( r_{\rm K} \) is the veiling, 
\( A_{\rm K} \) is the extinction, 
and \( d \) is the distance 
(\(K\) in the subscript denotes quantities measured in the $K$ band).
We adopted \( BC_{\rm K} = 1.33 \pm 0.03\,\mathrm{mag} \) from \citet{Mamajek}, corresponding to the effective temperature of \( T_{\rm eff} = 5919 \pm 93\,\mathrm{K} \) derived from our analysis (see Sect.~\ref{par:remove}).  
For the other parameters, we used \( m_{\rm K} = 8.73 \pm 0.01\,\mathrm{mag} \) from synthetic photometry and \( A_{\rm K} = 1.0 \pm 0.2\,\mathrm{mag} \).  
We converted the $A_{\rm V}$ values from Sect.~\ref{sec:lassALCALA} to $K$ band extinction using the \cite{Cardelli} law.  
With these values, we obtained \( M_{\rm bol} = -0.35 \pm 0.65 \,\mathrm{mag} \) and \( 0.58 \pm 0.32 \,\mathrm{mag} \) for the two distance estimates, \( d_{\rm BJ} \) and \( d_{\rm NGC} \), respectively.

From the bolometric magnitude, we derive the stellar luminosity using
\begin{align}
    \log\left(\frac{L_\star}{L_\odot}\right) = 0.4 \left(M_{\rm bol,\odot} - M_{\rm bol}\right),
\end{align}
where \( M_{\rm bol,\odot} = 4.74\,\mathrm{mag} \) is the solar bolometric luminosity. 
The resulting luminosities are approximately \( 110\,\rm L_\odot \) for \( d_{\rm BJ} \) and \( 45\,\rm L_\odot \) for \( d_{\rm NGC} \).
The exact numbers are presented in Table \ref{tab:stellprop}.
The stellar parameters were derived using the online tool provided by Lionel Siess\footnote{\url{http://www.astro.ulb.ac.be/~siess/pmwiki/pmwiki.php/WWWTools/HRDfind}}, which implements the pre-main sequence evolutionary models of \citet{Siess2000}.

We assumed a metallicity of $Z = 0.02$, but using different metallicity values did not significantly affect the results.
This method provides estimates for the stellar mass and radius, yielding $M_\star = 3.2-4.3 \rm \,M_\odot$ and $R_\star = 6-9.7 \rm \,R_\odot$, depending on the assumed distances.

Finally, using the stellar parameters and the accretion luminosities derived earlier from the empirical relationships and from the SED, we computed the mass accretion rate with the standard magnetospheric accretion formula
\begin{align}
    \dot{M}_{\rm acc} = \left(1 - \frac{R_\star}{R_{\rm in}}\right)^{-1} \frac{L_{\rm acc} R_\star}{G M_\star} = 1.25\,\frac{L_{\rm acc} R_\star}{G M_\star},
\end{align}
where \( G \) is the gravitational constant and the factor 1.25 assumes an inner disk truncation radius \( R_{\rm in} = 5 \rm R_\star \) \citep{Hartmann16}.

Figure~\ref{fig:mass} shows the extinction-corrected $\dot{M}_\mathrm{acc}$ for both distance estimates and both relationships. 
Table~\ref{tab:stellprop} presents the rates based on the SED and presents the stellar properties and  Table~\ref{tab:mass} lists the rates derived from individual tracers.

\begin{figure}
    \centering
    \includegraphics[width=\linewidth]{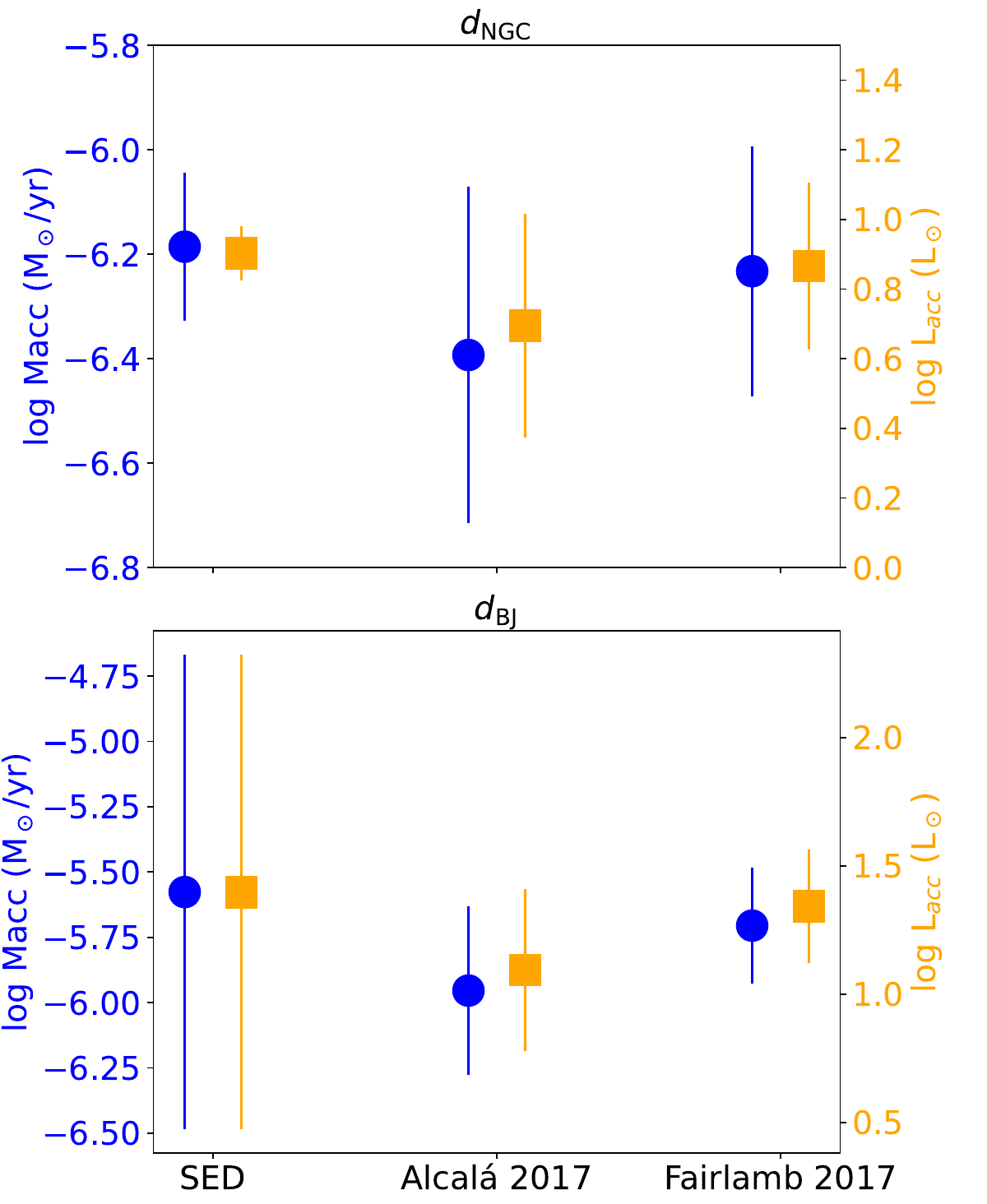}
    \caption{$L_{\rm acc}$ and $M_{\rm acc}$ based on SED, \cite{Alcala17} relationship, and on \cite{Fairlamb17} relationship. The orange squares show the $L_{\rm acc}$, while the blue dots show the corresponding $M_{\rm acc}$. The presented accretion properties from the two aforementioned relationships are the mean values of the individual measurements of each accretion tracer line summarized in Table~\ref{tab:minden}.}
    \label{fig:LaccMacc}
\end{figure}

\section{Discussion}\label{sec:discussion}
The results from Sect.~\ref{sec:lassALCALA} and Sect.~\ref{sec:parameter} are presented in Fig.~\ref{fig:LaccMacc}.
Points labeled \cite{Alcala17} and \cite{Fairlamb17} show the average \(L_{\rm acc}\) and \(M_{\rm acc}\) from accretion-tracing lines (see Sect.~\ref{sec:lassALCALA}).
The results for each line used in the averaging are in Table~\ref{tab:mass}.
Overall, the $L_{\rm acc}$ and $M_{\rm acc}$ values derived from the SED are consistent with those from the empirical relationships, supporting the conclusion that the observed brightening resulted from enhanced accretion.
We note that spectroscopic follow-up, including a quiescent spectrum and a corresponding quiescent $M_{\rm acc}$ determination, would provide an independent check. This would further support the observed changes in accretion.

Eruptive YSOs show a range of characteristics, but comparisons with similar objects help place Gaia20dsk in context. 
Gaia19bey, a MNor, is one such object, sharing characteristics with Gaia20dsk \citep{Hodapp2020}.
Both sources are embedded YSOs with similar classifications, namely: Gaia19bey is a Flat-spectrum source, while Gaia20dsk falls into the Class~I or FS category.
In terms of spectroscopic features, both objects exhibit strong \ion{H}{I} emission lines along with a prominent \ion{Ca}{II} IR triplet. 
Molecular hydrogen emission is also present in both spectra. 
Additionally, they occupy a similar region and exhibit comparable variations in the $J-H$ versus $H-K_{\rm s}$ color--color diagram (see Fig.~\ref{fig:ccd_2mass}).
During its 2019 outburst, Gaia19bey exhibited an accretion rate of $1.5^{+1.9}_{-0.7} \times 10^{-6}\, \mathrm{M_\odot}\,\mathrm{yr}^{-1}$. 
The mass accretion rates derived for Gaia20dsk are comparable, 
with $M_{\rm acc} = 5.4^{+4.1}_{-2.3} \times 10^{-7}\, \mathrm{M_\odot\,yr^{-1}}$ for $d_{\rm NGC}$ 
and $M_{\rm acc} = 1.8^{+2.1}_{-1.0} \times 10^{-6}\, \mathrm{M_\odot\,yr^{-1}}$ for $d_{\rm BJ}$.
    
Prior to late 2019, the star exhibited no changes in brightness since 2015, when the first Gaia measurements were taken.
However, since then, Gaia20dsk has experienced two major burst events and one brief burst.
During the first major burst, Gaia20dsk brightened by about 1.8 magnitudes in the $G$ band, with a rate of 0.15 magnitudes per month for approximately one-and-a-half years, and the entire burst event lasted for two years.
After the first burst, Gaia20dsk did not return to its previous brightness state.
It is possible that the enhanced accretion rate did not revert to the quiescent stage, but only decreased.
The duration of Gaia20dsk’s first major burst, of approximately two years, aligns with MNor-type variability timescales, namely, of $> 1.5$ years \citep{MNOR2017}.
\setlength{\parskip}{0pt}

The WISE light curve shows only weak variability, with changes less pronounced than those seen in the optical bands. 
Our data do not allow firm conclusions regarding the origin of this behavior in Gaia20dsk.
What we can conclude is that the WISE photometry of Gaia20dsk is not significantly affected by contamination from nearby sources, and that the NIR changes are not attributable to variable extinction, as supported by our color–color diagram analysis.
Similar cases of weak NIR variability compared to the optical have been reported in other eruptive YSOs, such as Gaia18cjb, a Class I object \citep{18cjb}, and EX Lupi during its 2022 outburst \citep{EXLupi}.
However, the physical origin of such weak IR responses may differ from source to source.

The NIR color light curve and color--color diagram show that Gaia20dsk became bluer during its burst events.
This behavior is typical for young stars experiencing outbursts, as they often become bluer, shifting their position toward the T Tauri locus in the diagram \citep[e.g.,][]{Lorenzetti12, Hodapp19}.
The changes observed in Gaia20dsk are similar to those seen in other well-known young eruptive stars, such as V1647 Ori \citep[e.g.,][]{V1647Ori}, Gaia19bey \citep{Hodapp2020} and V2775 Ori \citep{Garatti2011}.  
Based on the VIRAC and 2MASS photometry, we conclude that the brightening observed in 1998 was a real physical event and not the result of changing extinction. This is supported by the fact that the observed variations do not follow the extinction path and show a bluer trend similar to that seen during the outburst of V2775 Ori (see Fig.~\ref{fig:ccd_2mass}).
Additionally, Gaia20dsk's 1998 flux densities (2MASS) were significantly higher than in 2020 (synthetic photometry), exceeding the 3$\sigma$ level and indicating a stronger burst 22 years earlier.

The spectrum reveals a prominent H$\alpha$ emission along with the [\ion{S}{II}] lines at 6716~\AA{} and 6731~\AA. 
These features, taken together, typically indicate the presence of Herbig–Haro shocks \citep{Hodapp2020}. 
Furthermore, the \ion{Ca}{II} triplet lines are the strongest among the IR lines, a characteristic commonly observed in T~Tauri stars \citep[e.g.,][]{Muzerolle98}.
Under optically thin conditions, the expected intensity ratio of these three \ion{Ca}{II} lines is approximately 1:9:5, as reported by \citep{Azevedo2006} for CTTS.
In the case of Gaia20dsk, however, the observed ratio approaches unity, suggesting that the emission originates from a region that is a nearly optically thick.
This behavior parallels that seen in Gaia19bey, which has been identified as a MNor.
Notably, this star also displayed combined H$\alpha$ and [\ion{S}{II}] emission features.
Our spectrum also exhibits emission from H$_2$ at 2.12~\textmu m, which is associated with shocks caused by molecular outflows connected to the outburst event, which is an important characteristic of MNors \citep{MNOR2017}.

\section{Conclusion}
Our analysis of photometric and spectroscopic data indicates that Gaia20dsk has undergone an accretion outburst. Its properties do not clearly correspond to classical EXor or FUor categories; we propose that Gaia20dsk represents an intermediate case resembling that of MNors. Following the criteria defined by \citet{MNOR2017} for this subclass, we compared Gaia20dsk’s characteristics to these standards:

1) SEDs corresponding to Class I or FS sources, which Gaia20dsk also satisfies (see Sect.~\ref{sec:szintetikus}).

2) Outburst durations lasting longer than 1.5 years, but shorter than those of FUors. Gaia20dsk meets this criterion, with its first major burst lasting over two years (see Fig.~\ref{fig:light}).

3) Spectroscopic features typical of eruptive variables. This is also true for Gaia20dsk, which shows accretion tracer lines common in outbursting YSOs (see Sect.~\ref{sec:lines}). 
Although CO emission is absent, this does not contradict its classification, as CO lines may be suppressed by strong veiling or envelope emission \citep{MNOR2017}.

4) MNors typically exhibit H$_2$ emission at 2.122~\textmu m. This emission line is detected in our spectrum as well (see Fig.~\ref{fig:h2_line}).

Furthermore, while it was not set as a formal criterion, \citet{MNOR2017} found that MNors usually have a bolometric luminosity approximately one order of magnitude greater than their accretion luminosity. This feature is also observed in Gaia20dsk, where this ratio is roughly 0.1.
In addition, another characteristic of MNors is that they have a high bolometric luminosity in the range of hundreds to thousands of $L_\sun$, which holds for Gaia20dsk in the case of $d_{BJ}$ distance ($\sim 230-260 \rm \,L_\odot$) and close to this limit in $d_{\rm NGC}$ distance ($\sim 90-100 \rm \,L_\odot$).

Considering these criteria and the fact that our source shows similarities with such objects as V1647~Ori, Gaia19bey, and Gaia18cjb (three stars known to be intermediate between EXors and FUors), we conclude that Gaia20dsk is a member of the MNor-type YSos.

\begin{acknowledgements}
\\
\indent I dedicate this work to the memory of my beloved mother, Ibolya N\H{o}dl, whose love, guidance, and encouragement continue to inspire me. She was the one I loved the most, and her love, supports remain in my heart always.
\\
\indent We wish to thank the referee for a careful reading of this paper and for several major comments that significantly improved the conclusions.
Fernando Cruz-S\'aenz de Miera received financial support from the European Research Council (ERC) under the European Union’s Horizon 2020 research and innovation programme (ERC Starting Grant "Chemtrip", grant agreement No 949278).
Z.N. was supported by the J\'anos Bolyai Research Scholarship of the Hungarian Academy of Sciences.
Eleonora Fiorellino has been partially supported by project AYA2018-RTI-096188-B-I00 from the Spanish Agencia Estatal de Investigación and by Grant Agreement 101004719 of the EU project ORP.
We acknowledge the Hungarian National Research, Development and Innovation Office grant OTKA FK 146023. 
This work was also supported by the NKFIH NKKP grant ADVANCED 149943 and the NKFIH excellence grant TKP2021-NKTA-64. Project no.149943 has been implemented with the support provided by the Ministry of Culture and Innovation of Hungary from the National Research, Development and Innovation Fund, financed under the NKKP ADVANCED funding scheme. We acknowledge support from the ESA PRODEX contract nr. 4000132054.
Zs\'ofia M. Szab\'o acknowledges funding from a St. Leonards scholarship from the University of St. Andrews. Zs\'ofia M. Szab\'o of the International Max Planck Research School (IMPRS) for Astronomy and Astrophysics at the Universities of Bonn and Cologne.
Lukasz Wyrzykowski acknowledges Polish National Science Centre grant DAINA 2024/52/L/ST9/00210 and the European Union's Horizon Europe Research and Innovation programme ACME grant No 101131928.
\\
\indent This publication makes use of VOSA, developed under the Spanish Virtual Observatory project funded by MCIN/AEI/10.13039/501100011033/ through grant PID2020-112949GB-I00.
This work has made use of data from the Asteroid Terrestrial-impact Last Alert System project  primarily funded to search for near earth asteroids through NASA grants NN12AR55G, 80NSSC18K0284, and 80NSSC18K1575.
This research has made use of the NASA/IPAC Infrared Science Archive, which is funded by the National Aeronautics and Space Administration and operated by the California Institute of Technology.
We acknowledge ESA Gaia, DPAC and the Photometric Science Alerts Team.
This publication makes use of data products from the Near-Earth Object Wide-field Infrared Survey Explorer. It is funded by the National Aeronautics and Space Administration.
This research has made use of the SIMBAD database, operated at CDS, Strasbourg, France. 
This research has made use of the VizieR catalogue access tool, CDS, Strasbourg, France. 
\end{acknowledgements}
\bibpunct{(}{)}{;}{a}{}{,}
\bibliographystyle{aa}
\bibliography{bib.bib}

\begin{appendix}
\section{ATLAS data filtering} \label{app:filtering}
\begin{figure}[!ht]
    \centering
      \includegraphics[width=1\linewidth]{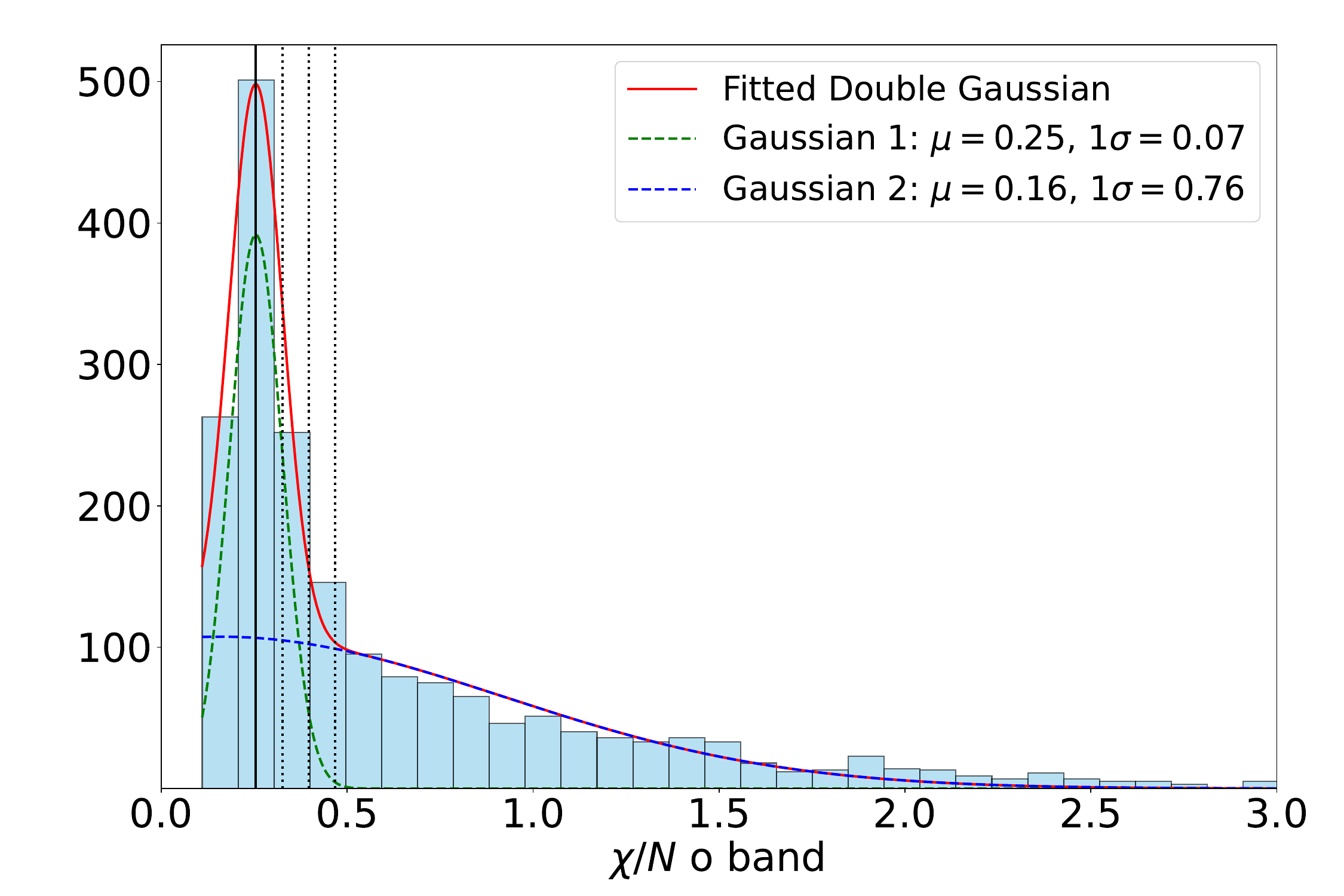}
      \includegraphics[width=\linewidth]{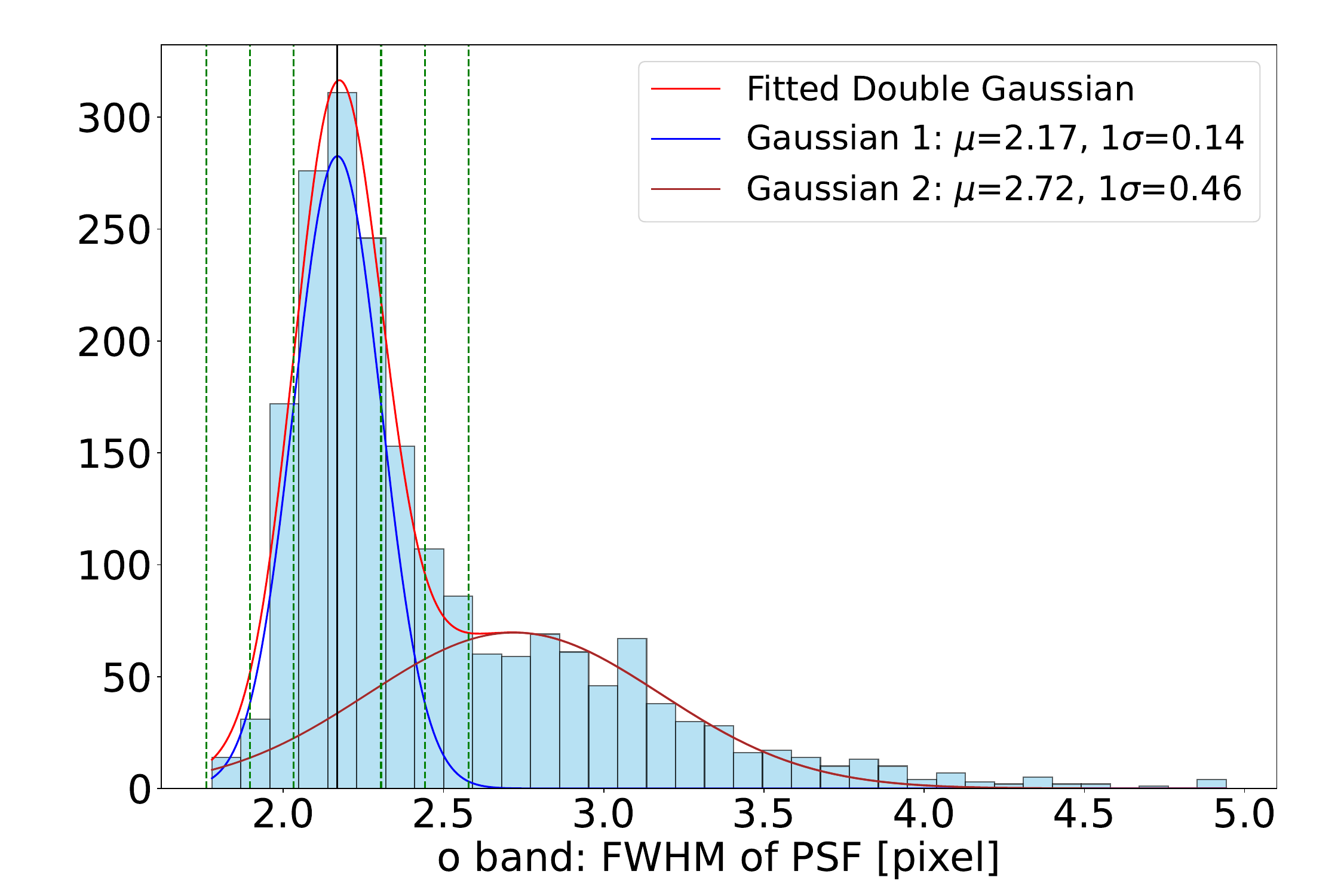}
      \caption{ATLAS filtering method for the $o$ band. The histograms of $\chi/N$ and $FWHM$ were fit with two Gaussian functions. The solid line indicates the mean of the Gaussian closest to the major peak of the distribution. The dashed lines represent this mean value plus/minus the $1\sigma$, $2\sigma$, and $3\sigma$ standard deviation.\label{fig:ATLAS_o_test}}
\end{figure}

To select reliable ATLAS photometry, we applied selection criteria based on the $\rm \chi/N$ value (which represents the reduced $\rm \chi^2$ PSF fitting of individual measurements), the S/N, and the FWHM of the PSF fit.
First, we examined the $\chi/N$ distribution using histograms (upper panel of Fig.~\ref{fig:ATLAS_o_test}), which was well approximated by a double Gaussian fit.
We focused on the narrow Gaussian component, i.e., the peak of the distribution, and set a filtering threshold at $\mu + 3\sigma$, where $\mu$ and $\sigma$ are the mean and standard deviation of the selected Gaussian.
Only data points satisfying $\chi/N < \mu + 3\sigma$ were retained.
A similar filtering approach was applied to the FWHM distribution (lower panel of Fig.~\ref{fig:ATLAS_o_test}).
Additionally, we retained only measurements with S/N $>$ 3.5, calculated using the fluxes and uncertainties in the ATLAS tables.
These three selection criteria were combined to refine the dataset before constructing the final light curve.

\section{WISE data analysis}\label{app:filtering2}

\subsection{Effect of contamination}
A nearby source, located to the north of Gaia20dsk and with 2MASS ID 17202041--3548117 
($\alpha_{2000} = 17^\mathrm{h}~20^\mathrm{m}~20.41^\mathrm{s}$, $\delta_{2000} = -35^\circ~48\arcmin~11.7\arcsec$), represents the strongest potential contaminant in the WISE bands.
To date, very little is known about this source; a SIMBAD query \citep{SIMBAD} using a 5\arcsec\ search radius returned no results, making it difficult to draw firm conclusions about its nature or behavior.
When contemporaneous measurements were available, the contaminating source was fainter by about 2\,mag in the $J$, $H$, and $K$ bands (based on 2MASS and VIRAC data), and by approximately 3\,mag in the Spitzer $I1$, $I2$ bands.

Here, the "contamination effect" refers to the fraction of flux in the WISE beam that originates from nearby source. 
Since no direct WISE measurements are available for the contaminant, we estimated its flux by converting contemporaneous Spitzer measurements into the WISE system using the color transformations of \citet{WISESPITZER}, and then into flux densities using the zeropoints from WISE documentation\footnote{\url{https://irsa.ipac.caltech.edu/data/WISE/docs/release/All-Sky/expsup/sec4_4h.html}}. 

Assuming total flux measured by WISE is given by

\begin{equation}
F_{\rm total,Wx} = F_{\rm G,Wx} + F_{\rm C,Wx}
,\end{equation}

where $F_{\rm G,Wx}$ and $F_{\rm C,Wx}$ denote the intrinsic fluxes of Gaia20dsk without the contamination and the contaminant, respectively. 
Using the values from Table~\ref{tab:sed}, $F_{\rm total,W1} = (1.14 \pm 0.06)\times10^{-14}$ and 
$F_{\rm total,W2} = (1.05 \pm 0.07)\times10^{-14}\ \rm erg\,s^{-1}\,cm^{-2}\,\AA^{-1}$, and the estimated contaminant fluxes of $F_{\rm C,W1} = (6.67 \pm 0.31)\times10^{-16}$ and $F_{\rm C,W2} = (4.80 \pm 0.19)\times10^{-16}\ \rm erg\,s^{-1}\,cm^{-2}\,\AA^{-1}$, 
we obtain

\[
F_{\rm G,W1} = (1.07 \pm 0.06) \times 10^{-14} \rm erg\,s^{-1}\,cm^{-2}\,\AA^{-1},
\]
\[
F_{\rm G,W2} = (1.00 \pm 0.07) \times 10^{-14}\ \rm erg\,s^{-1}\,cm^{-2}\,\AA^{-1}.
\]

The flux of Gaia20dsk is about 16 and 21 times higher than the contaminant in W1 and W2, respectively. The contaminant contribution can also be expressed in magnitude,

\begin{equation}
m_{\rm G} - m_{\rm total} = -2.5 \log_{10} \left( \frac{F_{\rm G}}{F_{\rm total}} \right)
,\end{equation}

which quantifies by how much the measured Gaia20dsk magnitude would be brightened due to the contaminant. 
For our data, we find $\Delta W1 \sim 0.08$\,mag and $\Delta W2 \sim 0.04$\,mag.
In other words, the presence of the contaminant would increase the observed brightness of Gaia20dsk by at most 0.08\,mag in W1 and 0.04\,mag in W2.

We found time-series photometry only in ATLAS database.
Applying the same filtering method as for Gaia20dsk, we determined that nearly all ATLAS points lie below the detection limit, and inspection of the corresponding images did not reveal the source.

Any potential contamination in the WISE measurements is at most comparable to, or smaller than, the measurement uncertainties, and therefore the contaminant is not expected to significantly affect Gaia20dsk’s observed photometry.
Additionally, WISE measures flux using PSF-fitting, which is less susceptible to contamination from nearby sources, and the contaminant only partially overlaps the beam.

\begin{figure}[ht]
    \centering
    \includegraphics[width=0.5\linewidth]{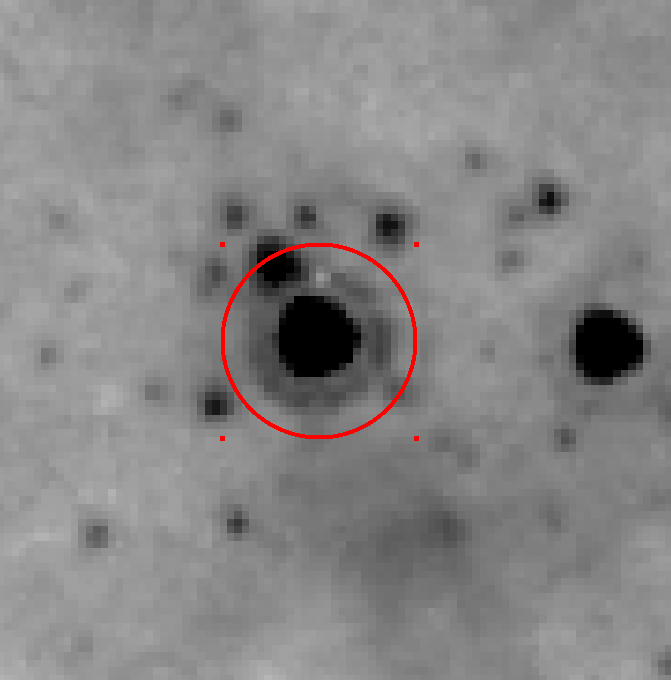}\\[2mm]
    \includegraphics[width=0.5\linewidth]{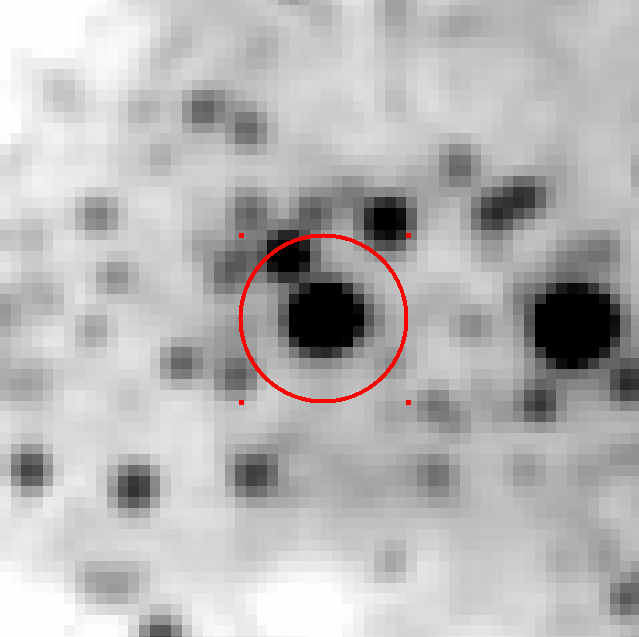}\\[2mm]
    \includegraphics[width=0.5\linewidth]{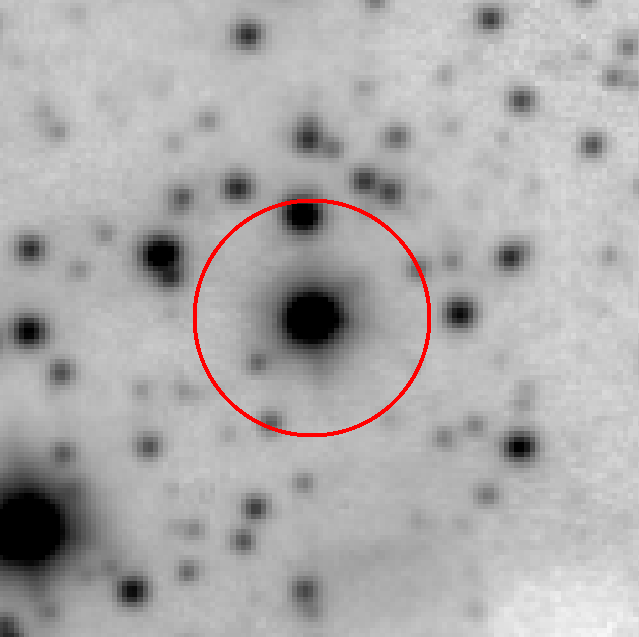}
    \caption{Images of Gaia20dsk as observed by Spitzer $I2$ filter (top), 2MASS $K_{\rm s}$ filter (middle), and VISTA $K_{\rm s}$ filter (bottom). Circles with radii of $1.3 \times \mathrm{FWHM}_{\mathrm{W2}}$ are overplotted to illustrate the WISE beam size.}
    \label{fig:wise}
\end{figure}

\subsection{Data filtering}
To select the reliable WISE photometry, we first discarded measurements where the flux was smaller than its uncertainty.
Afterward, we filtered out the measurements based on the following criteria:
\textsc{cc\_flags} is lowercase letter, \textsc{ph\_qual} is not equal to 'X', \textsc{w1rchi2}, and \textsc{w2rchi2} are less than 50; \textsc{qi\_fact} is nonzero, \textsc{saa\_sep} is higher than five, \textsc{moon\_masked} is not one.
The detailed description of these parameters is available from the database website\footnote{\url{https://wise2.ipac.caltech.edu/docs/release/neowise/expsup/sec2_1a.html}}.
Due to the brightening of Gaia20dsk being close to the saturation limits in each band, we applied a correction, which is available on the WISE website\footnotemark\footnotetext{\url{https://wise2.ipac.caltech.edu/docs/release/neowise/expsup/sec2_1civa.html}}.
Finally, we computed the median of multiple exposures of each season.

\section{Line profiles and gaussian fits of accretion tracers}\label{app:gaussians}
\begin{figure}[!htb]
   \centering
   \includegraphics[width=\linewidth]{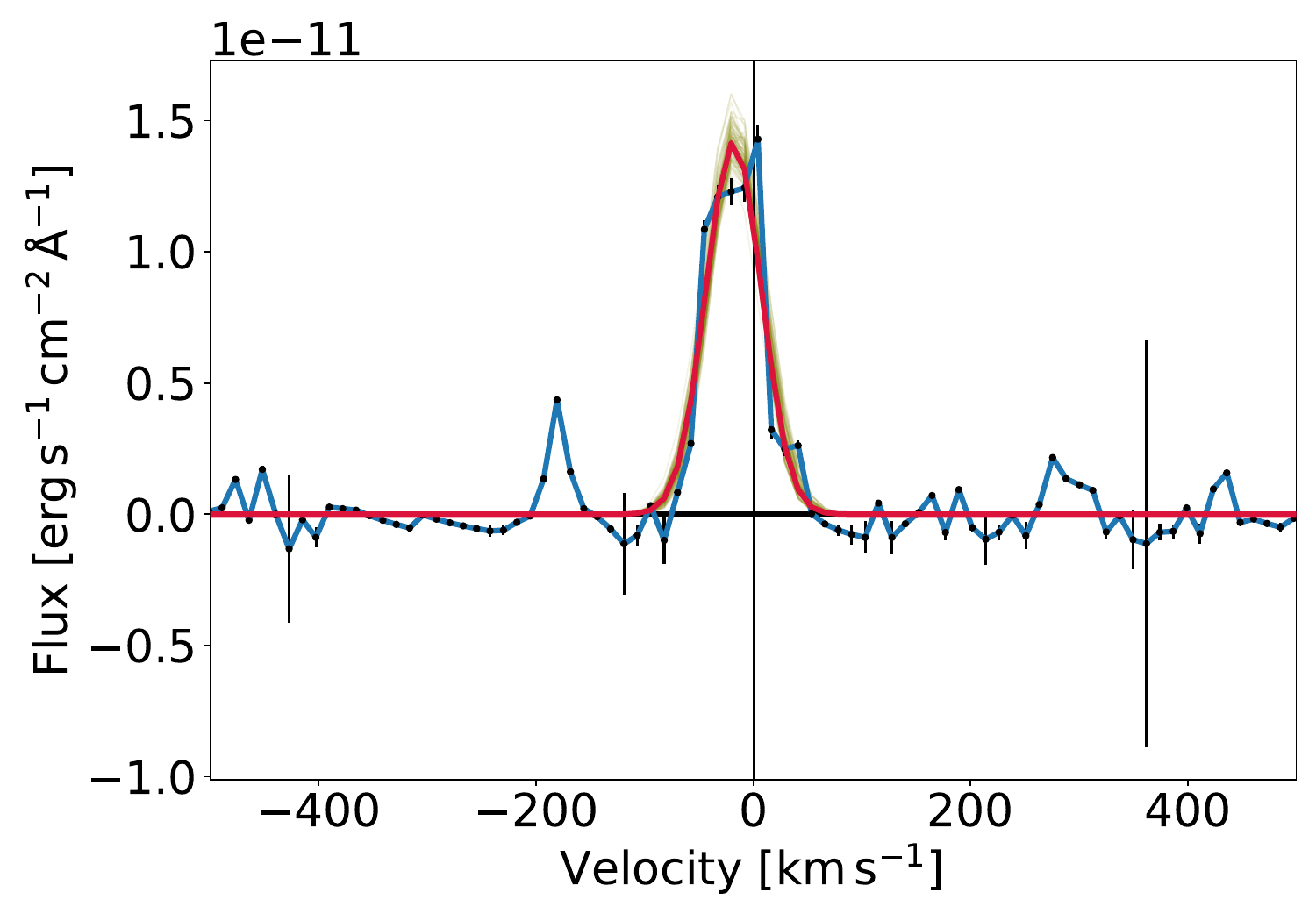}
   \caption{Extinction-corrected, photosphere-subtracted, and continuum-subtracted line profile of H$\mathrm{\beta}$ at 4861 \AA{}. The red curve represents the best fit of the Gaussian(s), while several individual samples showing the range of uncertainties.\label{app:hbeta}}
\end{figure}

\begin{figure}[!htb]
   \centering
   \includegraphics[width=\linewidth]{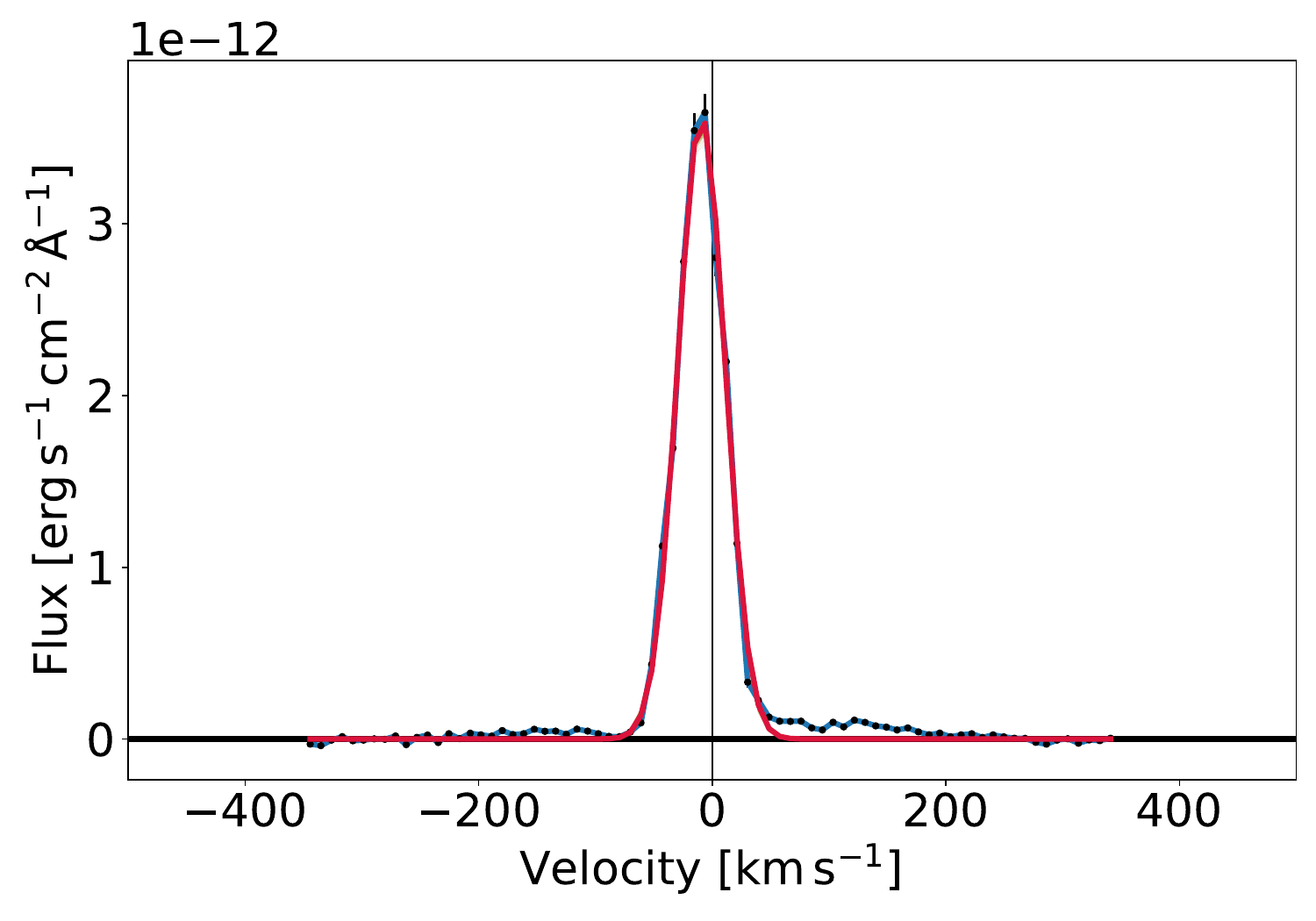}
   \caption{Same as Fig.~\ref{app:hbeta}, but for H$\mathrm{\alpha}$ at 6563 \AA{}.\label{app:halfa}}
\end{figure}

\begin{figure}[!htb]
   \centering
   \includegraphics[width=\linewidth]{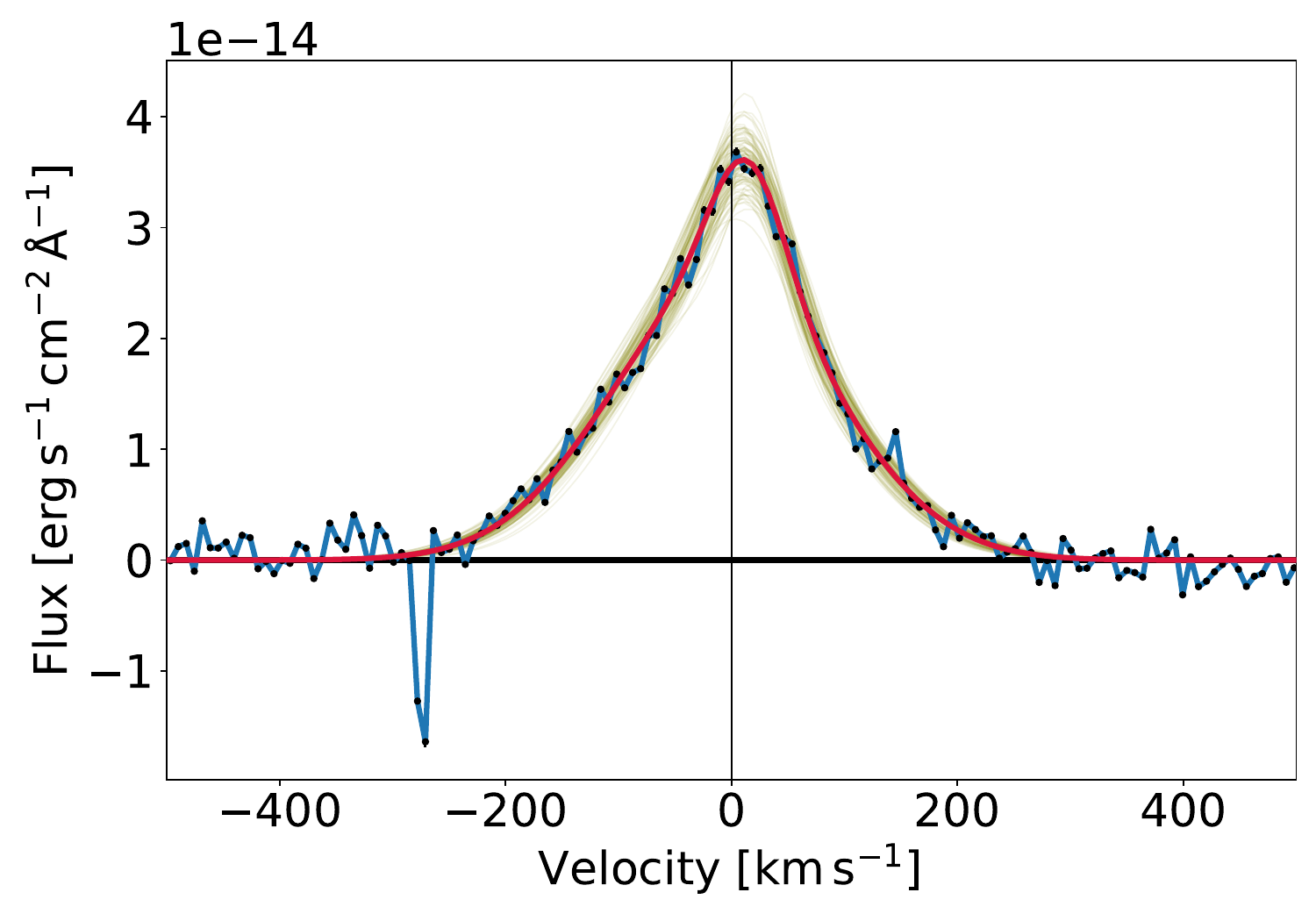}
   \caption{Same as Fig.~\ref{app:hbeta}, but for \ion{Ca}{II} at 8498 \AA{}.\label{app:caII1}}
\end{figure}

\begin{figure}[!htb]
   \centering
   \includegraphics[width=\linewidth]{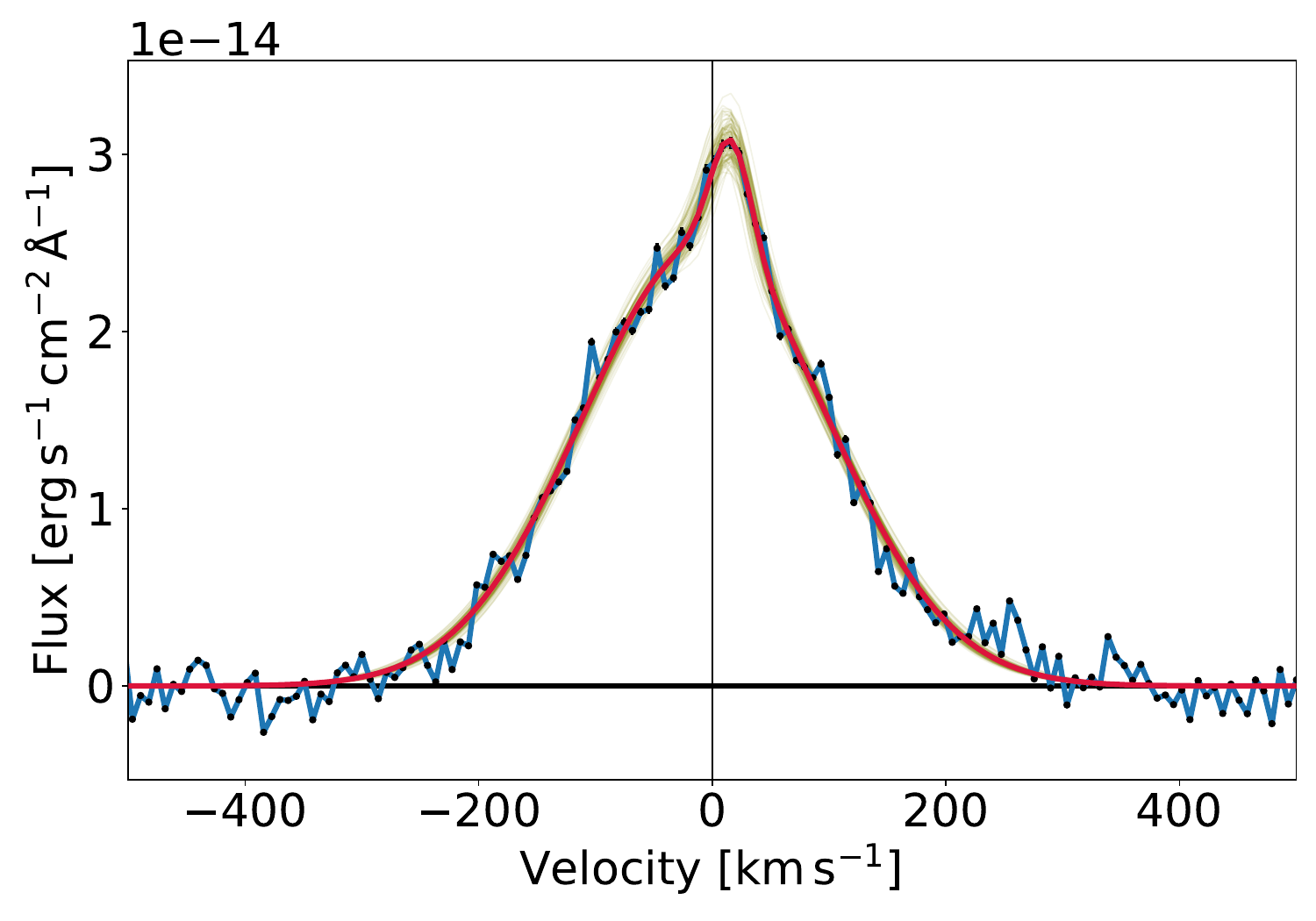}
   \caption{Same as Fig.~\ref{app:hbeta}, but for \ion{Ca}{II} at 8542 \AA{}.\label{app:caII2}}
\end{figure}

\begin{figure}[!htb]
   \centering
   \includegraphics[width=\linewidth]{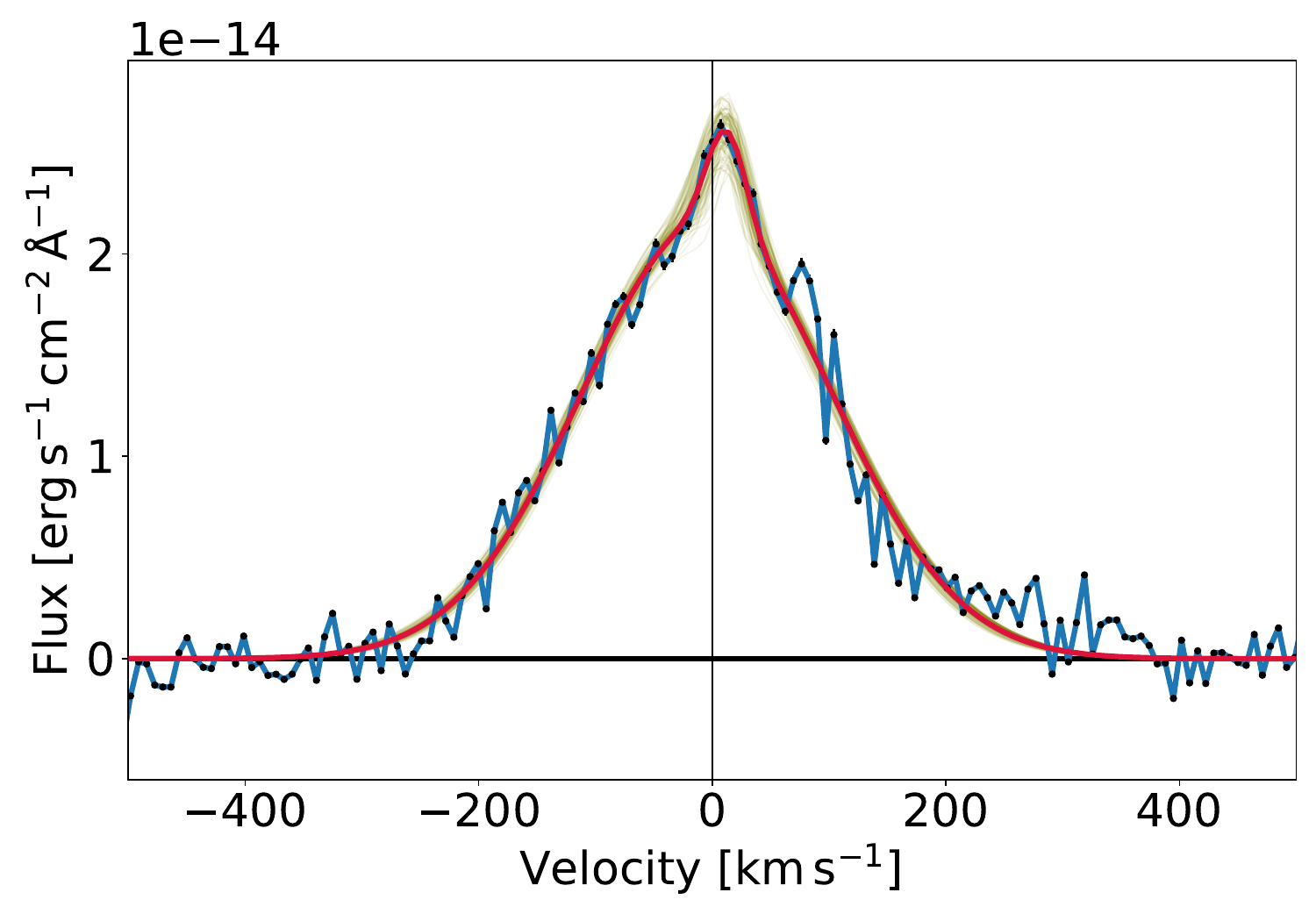}
   \caption{Same as Fig.~\ref{app:hbeta}, but for \ion{Ca}{II} at 8662 \AA{}.\label{app:caII3}}
\end{figure}

\begin{figure}[!htb]
   \centering
   \includegraphics[width=\linewidth]{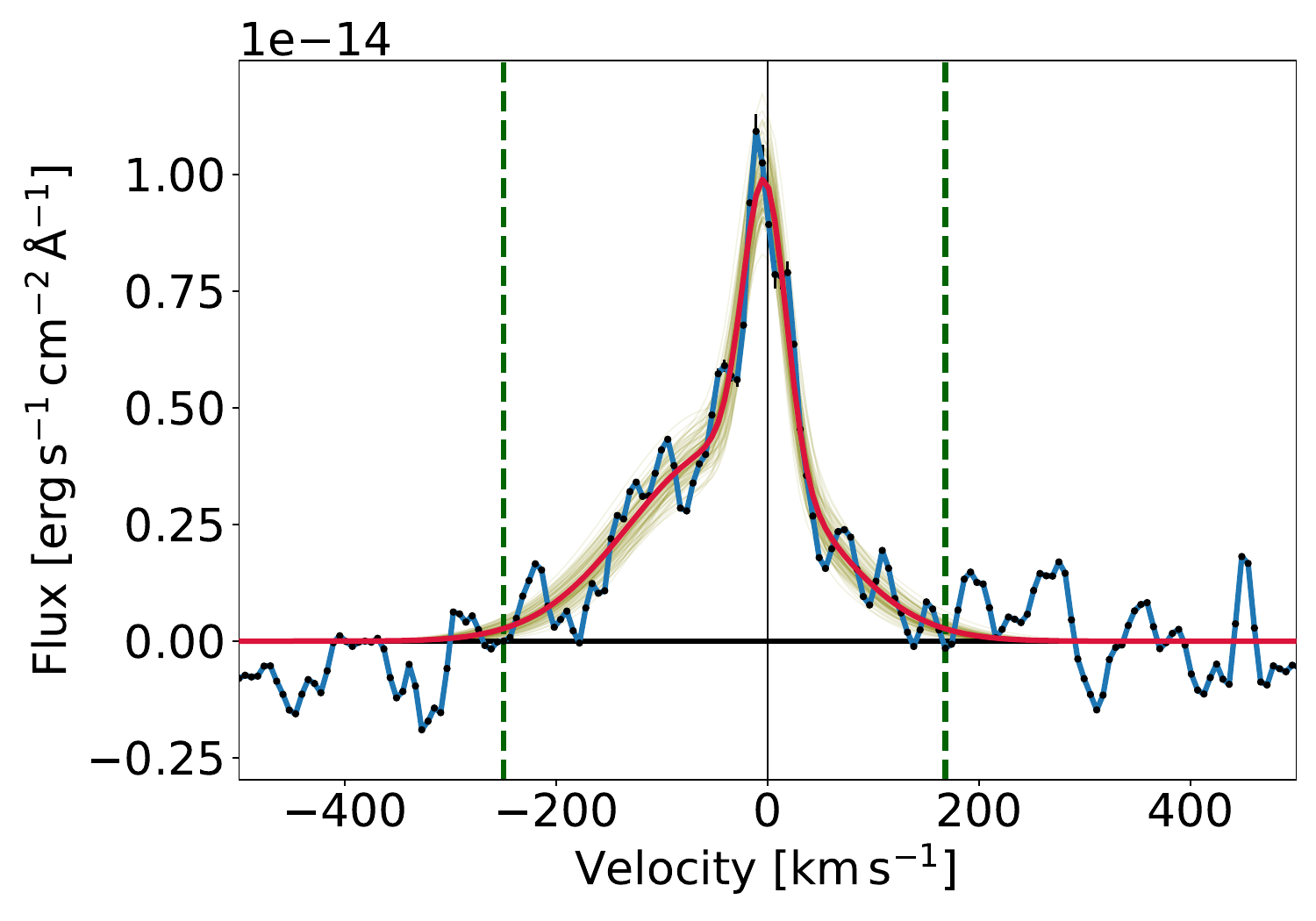}
   \caption{Similar to Fig.~\ref{app:hbeta}, but for Pa$\mathrm{\delta}$ at 10\,049 \AA{}.The green vertical dashed lines indicate the boundaries within which we calculated the line flux.\label{app:padelta}}
\end{figure}

\begin{figure}[!htb]
   \centering
   \includegraphics[width=\linewidth]{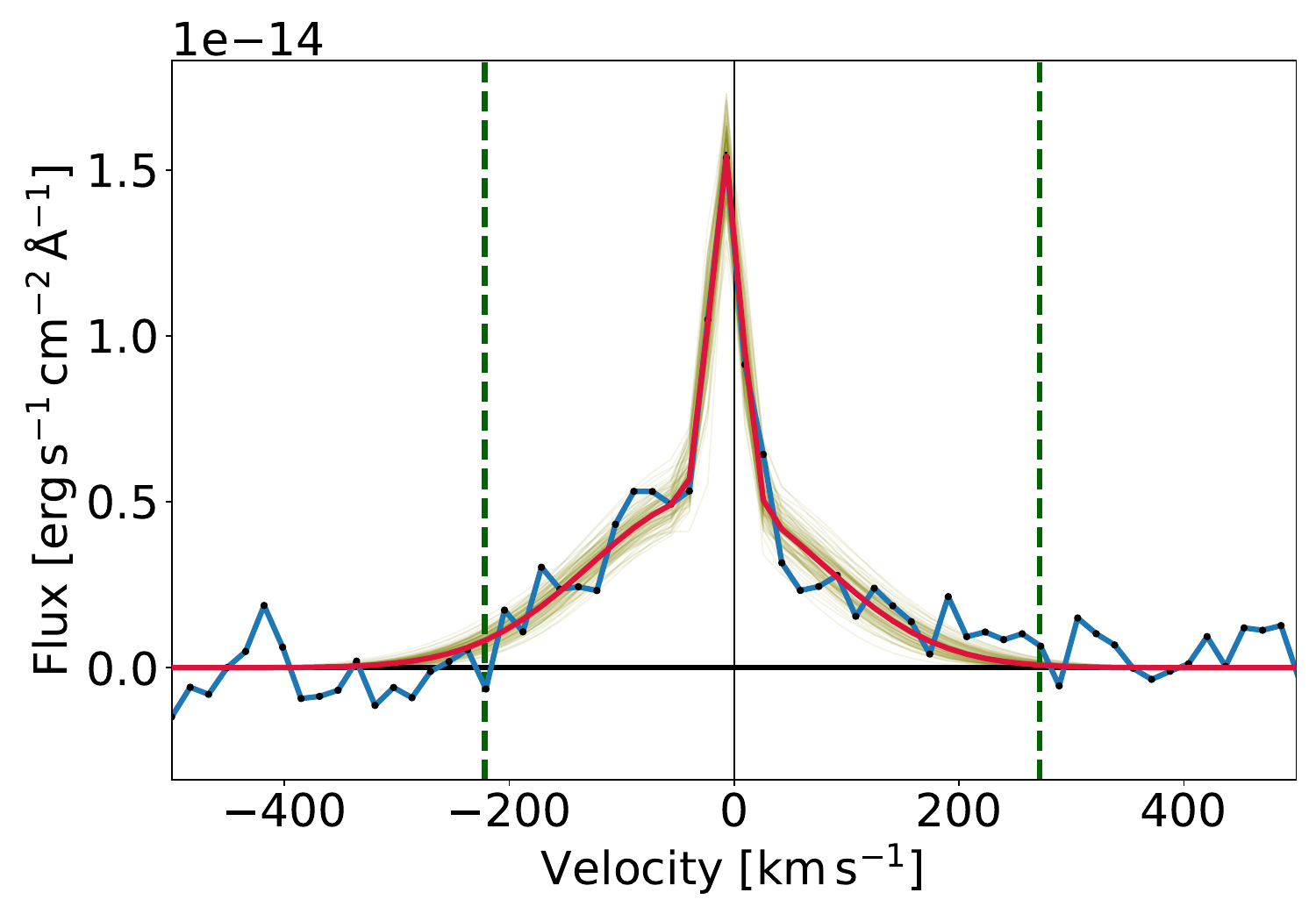}
   \caption{Same as Fig.~\ref{app:padelta}, but for Pa$\mathrm{\gamma}$ at 10\,938 \AA{}.\label{app:pagamma}}
\end{figure}

\begin{figure}[!htb]
   \centering
   \includegraphics[width=\linewidth]{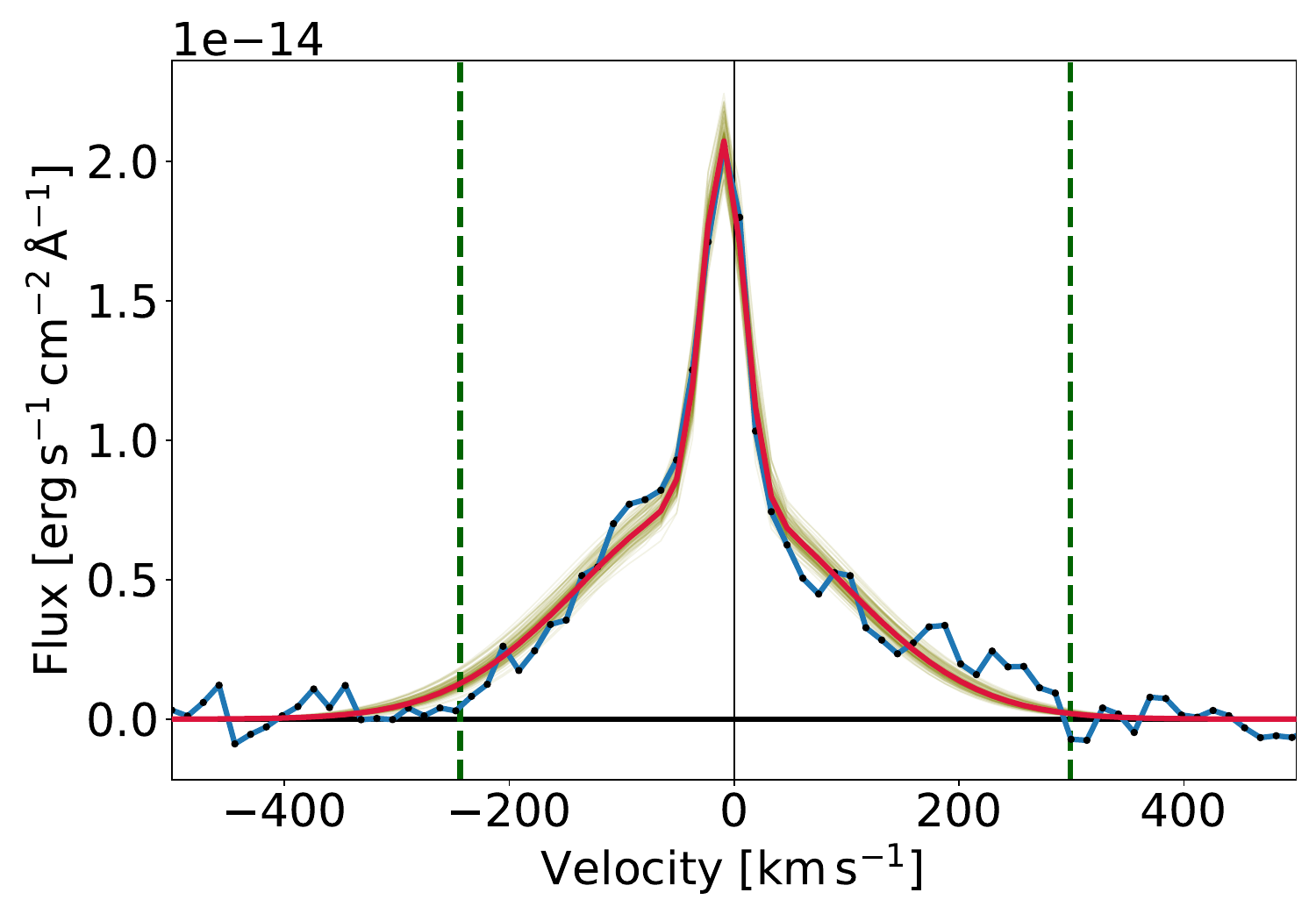}
   \caption{Same as Fig.~\ref{app:padelta}, but for Pa$\mathrm{\beta}$ at 12\,818 \AA{}.\label{app:pabeta}}
\end{figure}

\begin{figure}[!htb]
   \centering
   \includegraphics[width=\linewidth]{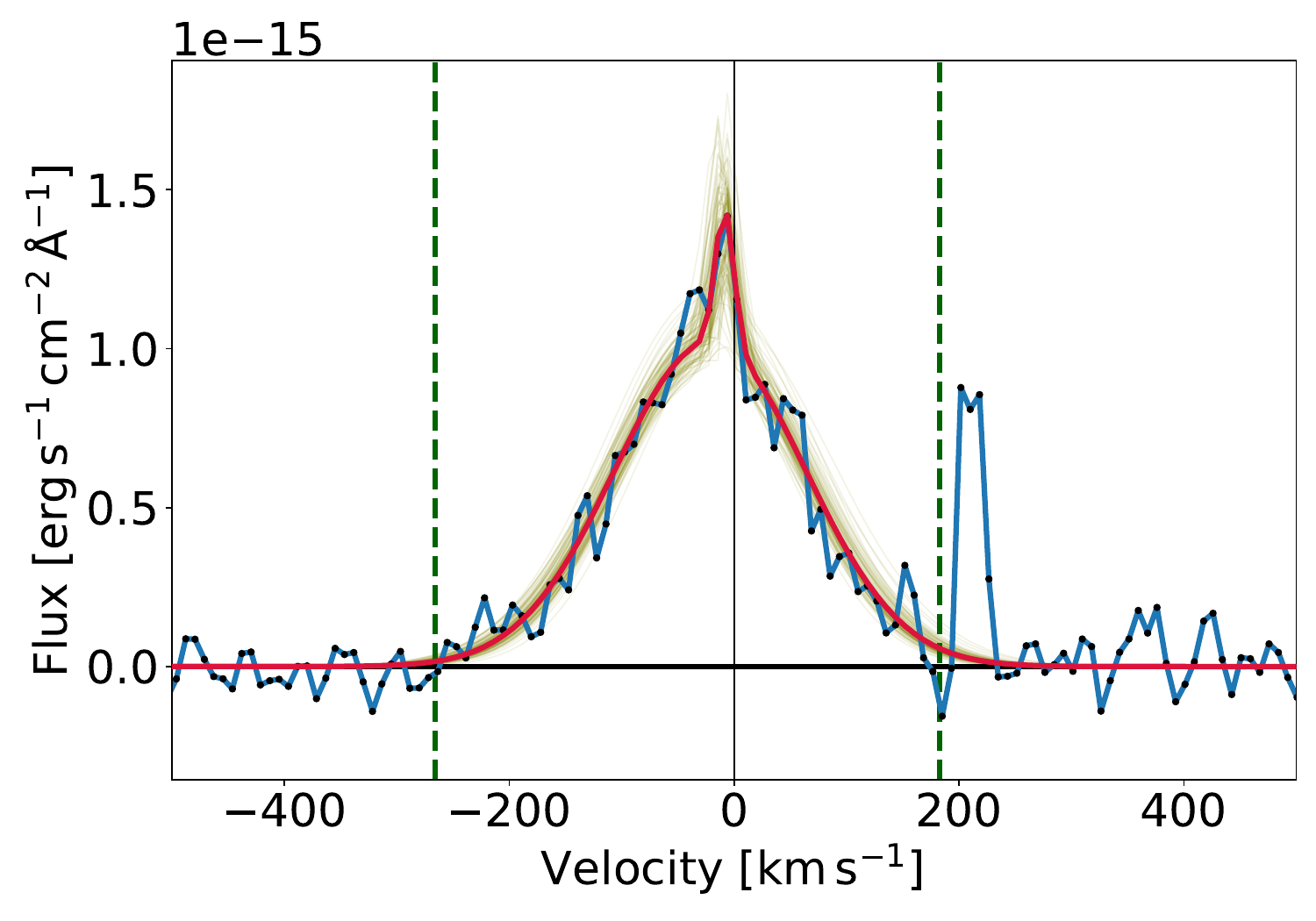}
   \caption{Same as Fig.~\ref{app:padelta}, but for Br$\mathrm{\gamma}$ at 21\,655 \AA{}.\label{app:brgamma}}
\end{figure}

\begin{figure}[!htb]
   \centering
   \includegraphics[width=\linewidth]{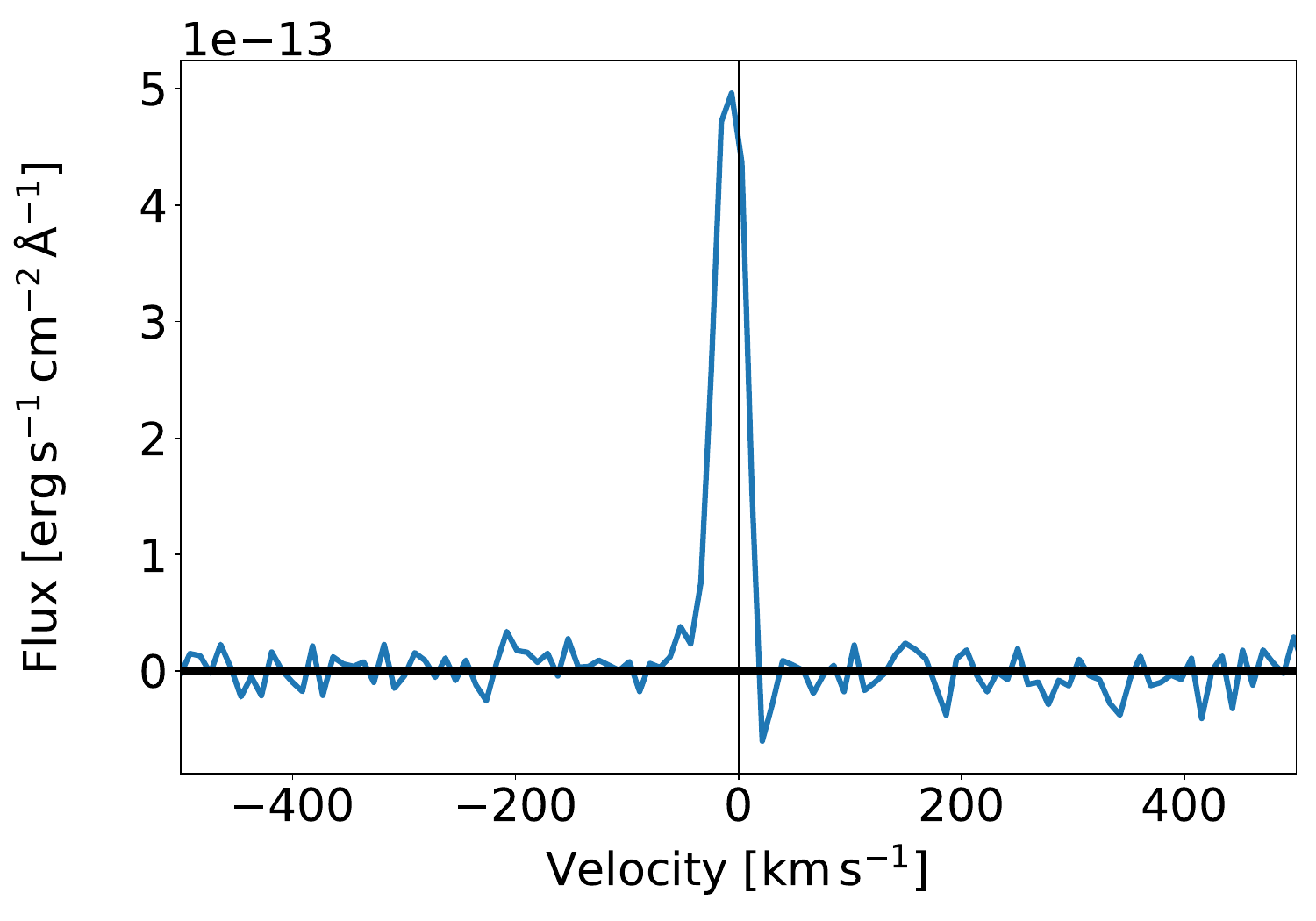}
   \caption{Extinction-corrected, photosphere-subtracted, continuum-subtracted line profile of [\ion{N}{II}] at 6548\text{ \text{\AA{}}}.\label{fig:nii_2_line}}
\end{figure}

\begin{figure}[!htb]
   \centering
   \includegraphics[width=\linewidth]{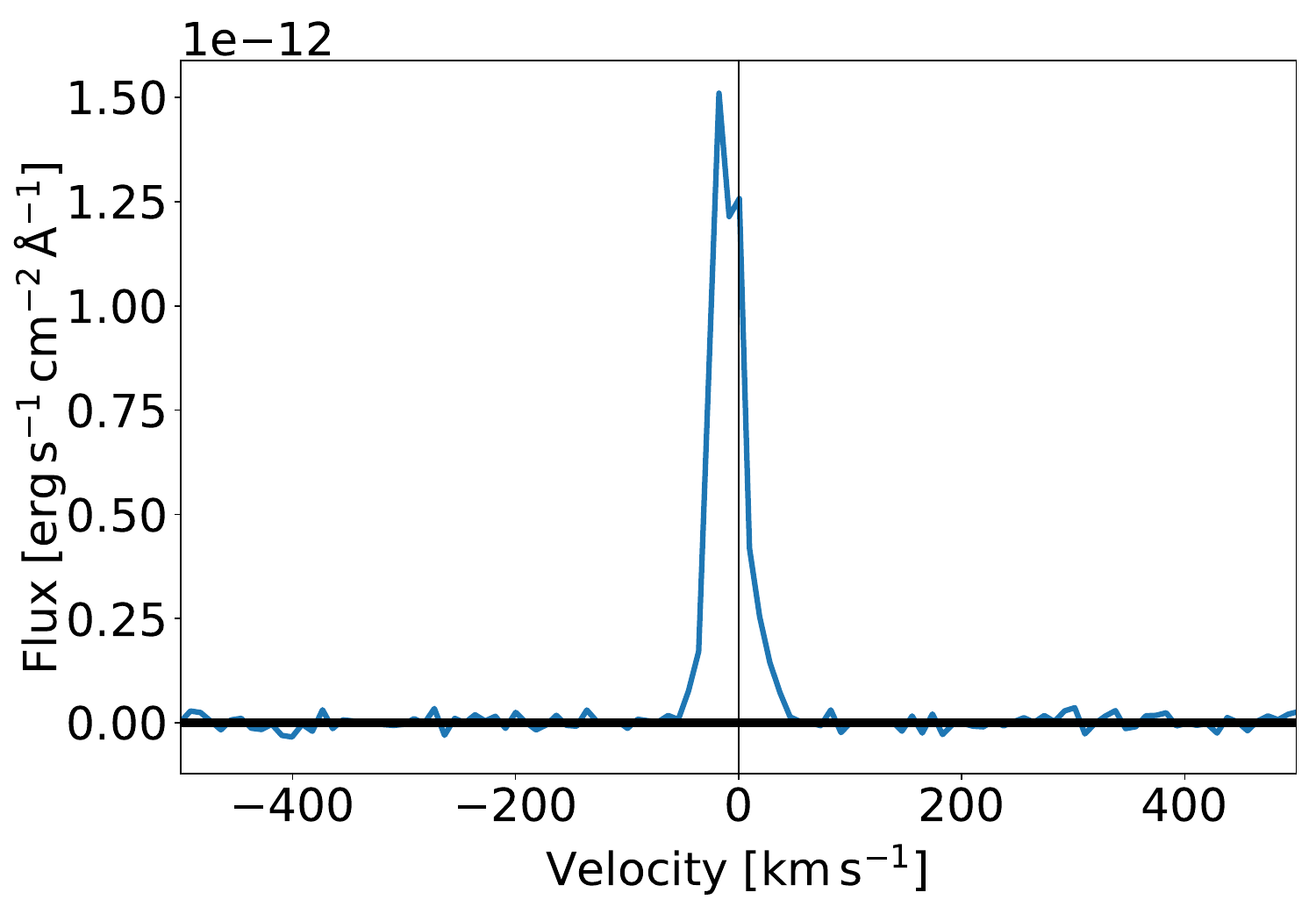}
   \caption{Same as \ref{fig:nii_2_line}, but for [\ion{N}{II}] at 6583\text{ \text{\AA{}}}.\label{fig:nii_1_line}}
\end{figure}

\begin{figure}[!htb]
   \centering
   \includegraphics[width=\linewidth]{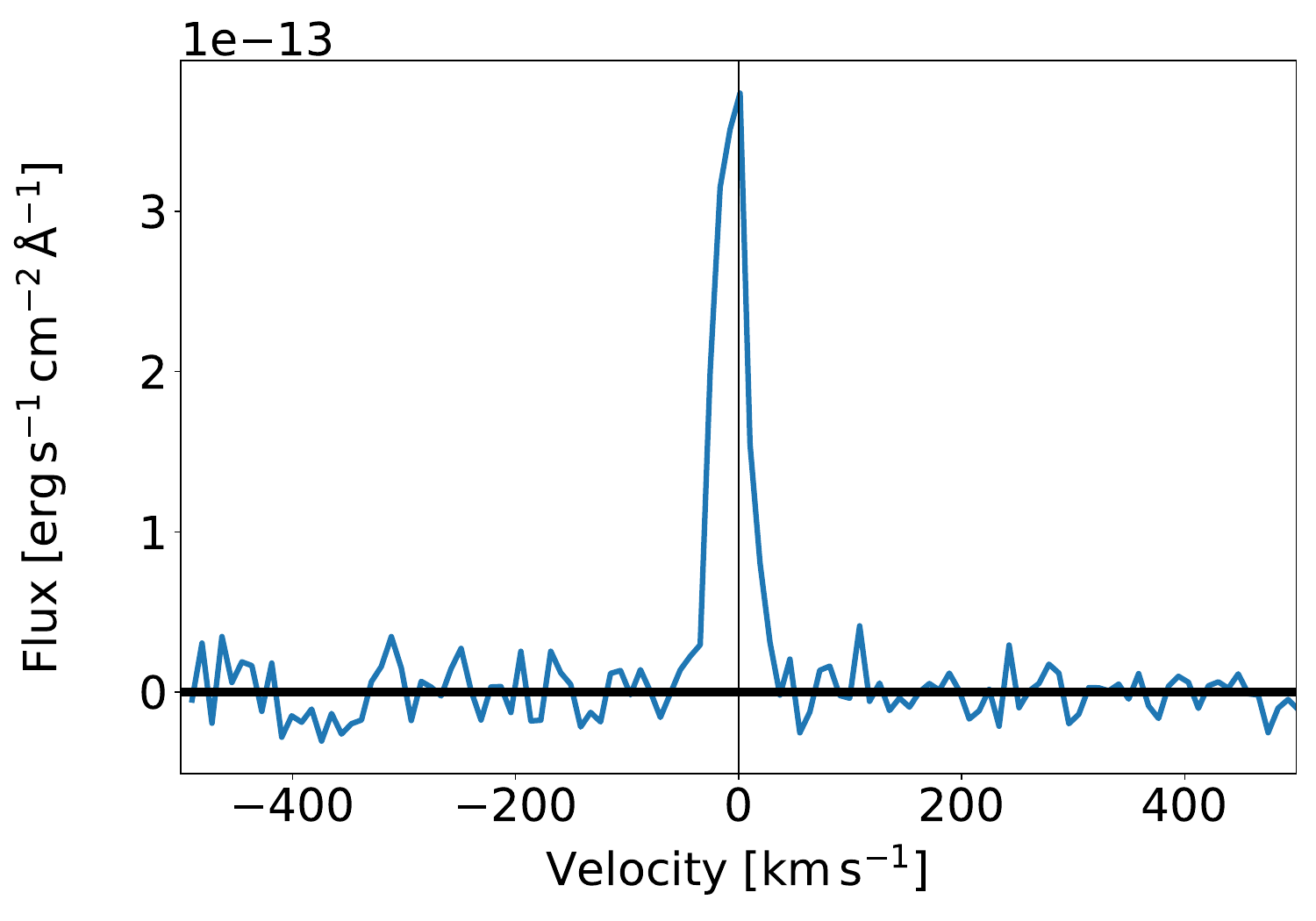}
   \caption{Same as \ref{fig:nii_2_line}, but for [\ion{S}{II}] at 6716\text{ \text{\AA{}}}.\label{fig:sii_1_line}}
\end{figure}

\begin{figure}[!htb]
   \centering
   \includegraphics[width=\linewidth]{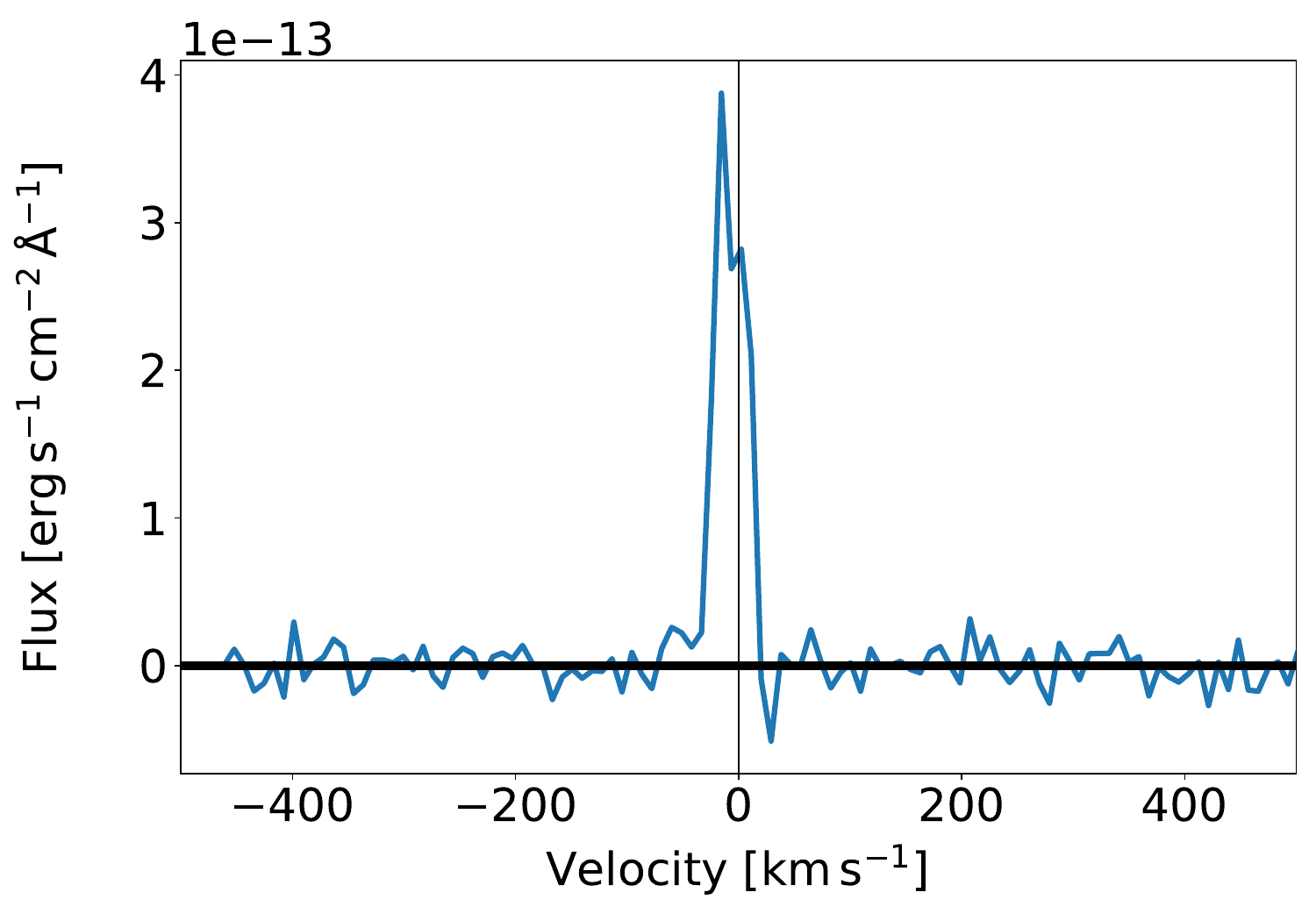}
   \caption{Same as \ref{fig:nii_2_line}, but for of [\ion{S}{II}] at 6731\text{ \text{\AA{}}}.\label{fig:sii_2_line}}
\end{figure}

\begin{figure}[!htb]
   \centering
   \includegraphics[width=\linewidth]{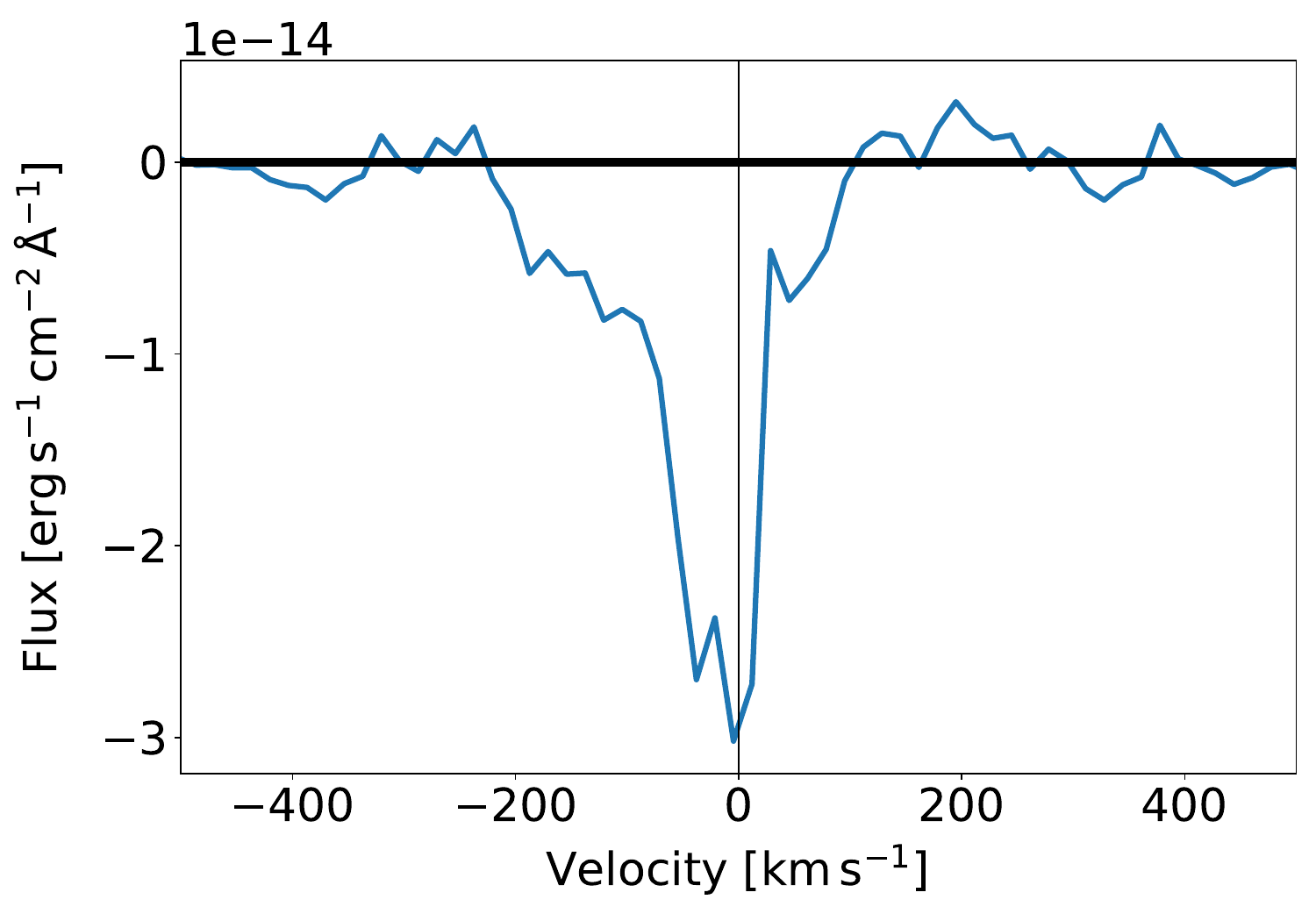}
   \caption{Same as \ref{fig:nii_2_line}, but for \ion{He}{I} at 10\,830\text{ \text{\AA{}}}.
   \label{fig:hei_line}}
\end{figure}

\begin{figure}[!htb]
   \centering
   \includegraphics[width=\linewidth]{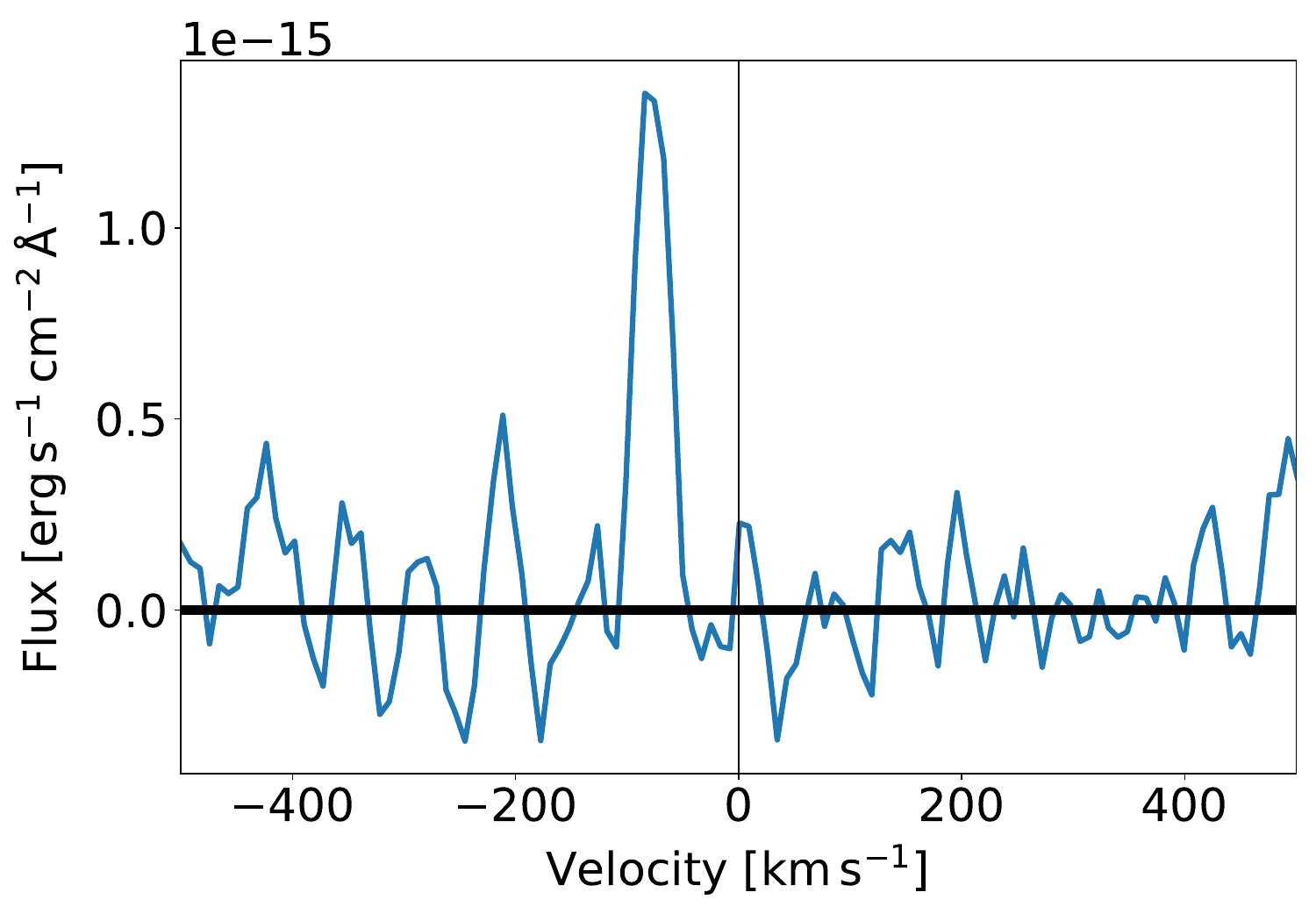}
   \caption{Same as \ref{fig:nii_2_line}, but for H$_2$ at 21\,218\text{ \text{\AA{}}}.\label{fig:h2_line}}
\end{figure}

\clearpage

\section{Flux densities used for the SED}\label{sec:sed_flux}
\begin{table}[!htb]
\centering
\caption{Archival and synthetic photometry of Gaia20dsk.\label{tab:sed}}
\begin{tabular}{lccc}
\hline
Filter & Wv\tablefootmark{a} & Flux & Mag. \\ 
 & [\textmu m] & [erg\,s$^{-1}$\,cm$^{-2}$\,\AA{}$^{-1}$] & [mag] \\ 
\hline
Gaia $G_{\rm BP}$ & $0.50$ & $(3.08\pm0.24)\times 10^{-17}$ & $20.31\pm0.09$ \\
Gaia $G$ & $0.59$ & $(4.90\pm0.06)\times 10^{-17}$ & $19.27\pm0.01$ \\
VPHAS+ $r$ & $0.62$ & $(2.63\pm0.14)\times 10^{-17}$ & $19.89\pm0.06$ \\
VPHAS+ $I$ & $0.75$ & $(5.68\pm0.15)\times 10^{-17}$ & $18.45\pm0.03$ \\
Gaia $G_{\rm RP}$ & $0.77$ & $(1.33\pm0.03)\times 10^{-16}$ & $17.43\pm0.03$ \\
X-S $I$\tablefootmark{b} & $0.79$ & $(4.18\pm0.19)\times 10^{-16}$ & $16.13\pm0.05$ \\
DENIS $I$\tablefootmark{c} & $0.79$ & $(8.51\pm0.31)\times 10^{-16}$ & $15.36\pm0.04$ \\
DENIS $I$\tablefootmark{d} & $0.79$ & $(1.09\pm0.04)\times 10^{-15}$ & $15.10\pm0.04$ \\
VIRAC $Z$ & $0.88$ & $(1.86\pm0.02)\times 10^{-16}$ & $16.69\pm0.01$ \\
VIRAC $Y$ & $1.02$ & $(3.42\pm0.03)\times 10^{-16}$ & $15.61\pm0.01$ \\
DENIS $J$\tablefootmark{c} & $1.22$ & $(5.59\pm0.35)\times 10^{-15}$ & $11.89\pm0.07$ \\
DENIS $J$\tablefootmark{d} & $1.22$ & $(7.32\pm0.46)\times 10^{-15}$ & $11.60\pm0.07$ \\
2MASS $J$ & $1.24$ & $(6.84\pm0.15)\times 10^{-15}$ & $11.66\pm0.02$ \\
X-S $J$\tablefootmark{b} & $1.24$ & $(3.92\pm0.06)\times 10^{-15}$ & $12.26\pm0.02$ \\
VIRAC $J$ & $1.25$ & $(6.65\pm0.06)\times 10^{-16}$ & $14.13\pm0.01$ \\
VIRAC $H$ & $1.63$ & $(5.98\pm0.05)\times 10^{-15}$ & $10.71\pm0.01$ \\
2MASS $H$ & $1.66$ & $(1.18\pm0.02)\times 10^{-14}$ & $9.97\pm0.02$ \\
X-S $H$\tablefootmark{b} & $1.66$ & $(9.01\pm0.02)\times 10^{-15}$ & $10.26\pm0.01$ \\
VIRAC $K_{\rm s}$ & $2.14$ & $(1.23\pm0.05)\times 10^{-14}$ & $8.88\pm0.05$ \\
DENIS $K_{\rm s}$\tablefootmark{c} & $2.15$ & $(1.68\pm0.08)\times 10^{-14}$ & $8.53\pm0.05$ \\
DENIS $K_{\rm s}$\tablefootmark{d} & $2.15$ & $(1.42\pm0.09)\times 10^{-14}$ & $8.71\pm0.07$ \\
2MASS $K_{\rm s}$ & $2.16$ & $(1.39\pm0.02)\times 10^{-14}$ & $8.73\pm0.02$ \\
X-S $K_{\rm s}$\tablefootmark{b} & $2.16$ & $(1.40\pm0.01)\times 10^{-14}$ & $8.73\pm0.01$ \\
WISE $W1$ & $3.35$ & $(1.14\pm0.06)\times 10^{-14}$ & $7.15\pm0.06$ \\
Spitzer $I1$ & $3.51$ & $(1.18\pm0.04)\times 10^{-14}$ & $6.89\pm0.04$ \\
Spitzer $I2$ & $4.44$ & $(8.97\pm0.51)\times 10^{-15}$ & $6.21\pm0.06$ \\
WISE $W2$ & $4.60$ & $(1.05\pm0.07)\times 10^{-14}$ & $5.92\pm0.08$ \\
Spitzer $I3$ & $5.63$ & $(6.58\pm0.13)\times 10^{-15}$ & $5.54\pm0.02$ \\
Spitzer $I4$ & $7.59$ & $(5.28\pm0.11)\times 10^{-15}$ & $4.47\pm0.02$ \\
\hline
\end{tabular}
\tablefoot{
\tablefoottext{a}{Effective wavelength of the photometry band}
\tablefoottext{b}{Synthetic photometry from the X-SHOOTER spectrum.}
\tablefoottext{c}{From 1998 data.}
\tablefoottext{d}{From 1999 data.}}
\end{table}

\clearpage
\onecolumn
\section{Stellar parmeters and accretion properties}

\begin{figure*}[!h]
    \centering
    \begin{subfigure}[b]{0.48\linewidth}
        \includegraphics[width=\linewidth]{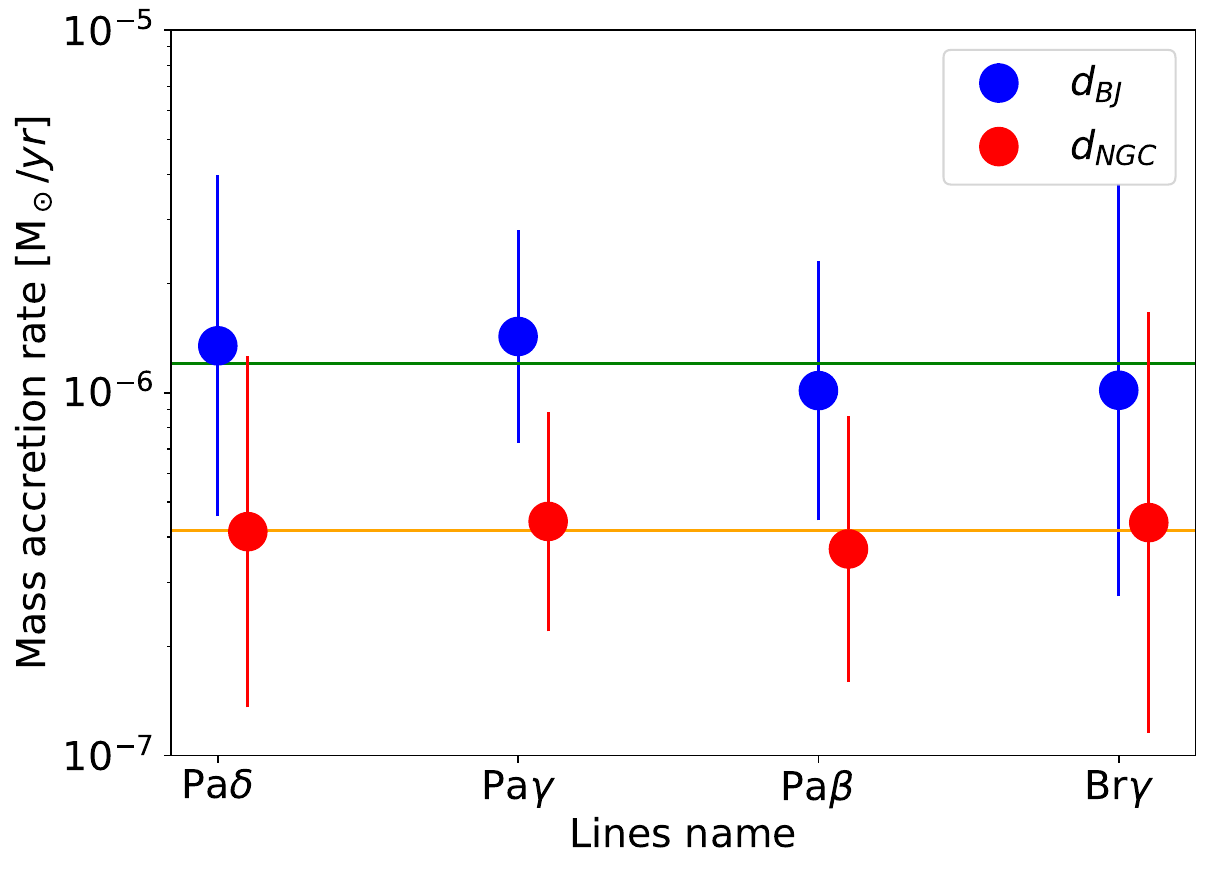}
        \label{fig:alcala}
    \end{subfigure}
    \hfill
    \begin{subfigure}[b]{0.48\linewidth}
        \includegraphics[width=\linewidth]{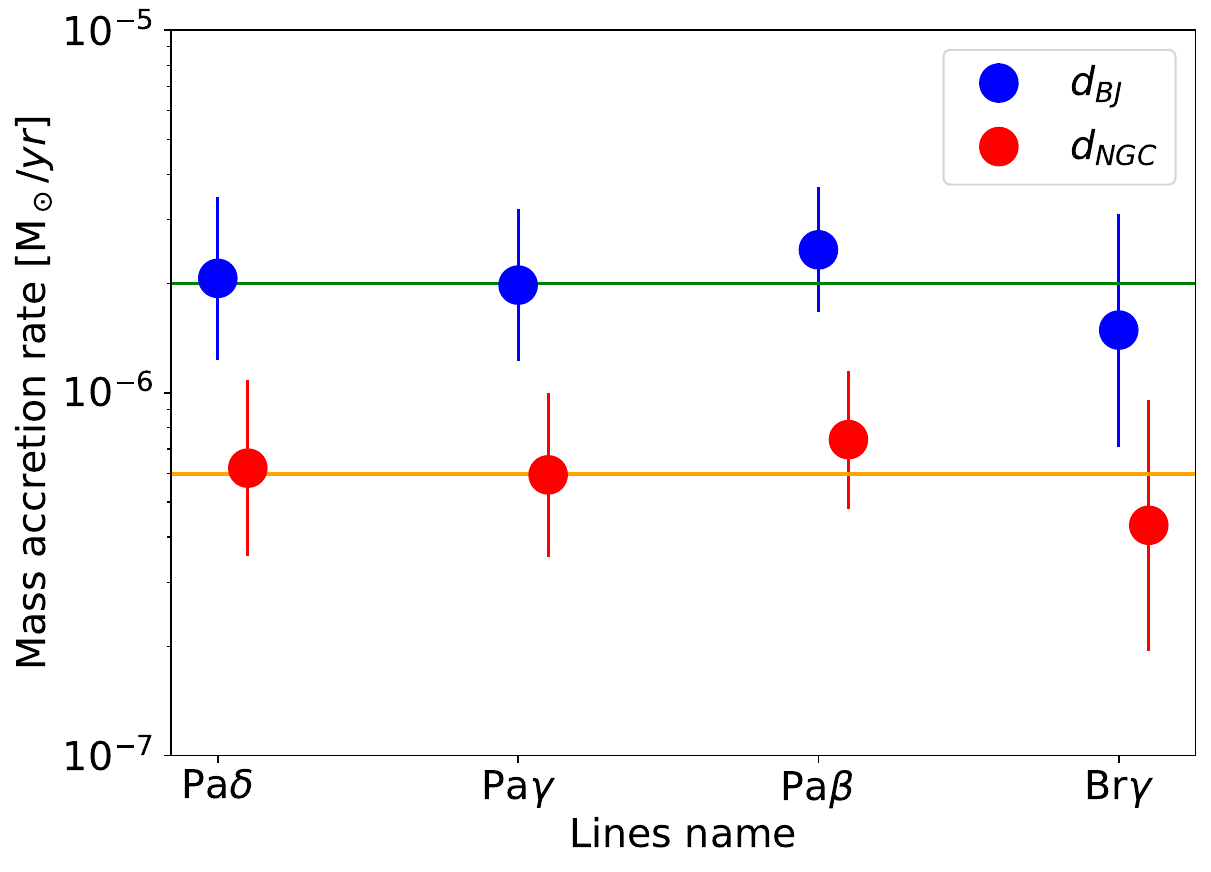}
        \label{fig:fairlamb}
    \end{subfigure}
    \caption{
    Mass accretion rate of Gaia20dsk per different accretion-tracing line and considering two different distances. The spectral lines are in order of ascending wavelength. 
    The horizontal green and orange lines show the mean value of the calculated mass accretion rate considering $d_{\mathrm{BJ}}$ and $d_{\mathrm{NGC}}$, respectively. The left panel shows the result based on \citet{Alcala17} coefficient, while the right panel shows the result based on the \cite{Fairlamb17} coeeficient. 
    }
    \label{fig:mass}
\end{figure*}

\end{appendix}

\end{document}